\documentclass[aps,twocolumn,amsmath,amsfonts,amssymb,nofootinbib,eqsecnum,%
  secnumarabic,notitlepage,superscriptaddress]{revtex4-2}
\usepackage[utf8]{inputenc}
\usepackage{amsmath,amsfonts,amssymb,graphicx,subfigure}
\usepackage{slashed,color}
\usepackage{multirow}
\usepackage{lineno}
\usepackage[svgnames]{xcolor}
\usepackage{hyperref} 
\hypersetup{colorlinks=true, linkcolor=blue, citecolor=ForestGreen,
filecolor=magenta, urlcolor=ForestGreen,
  pdftitle={Helicity inversion in Majorana neutrino decays, arXiv:2008.01092},pdfauthor={R. Ruiz},
  pdfsubject={Subject},pdfcreator={Creator},pdfproducer={Producer},
  pdfkeywords={Majorana Neutrinos}{Large Hadron Collider}{Spin Correlation}{Helicity Inversion}
}
\newcommand{\lnc}{{ L}}
\newcommand{\lnv}{{ \not\!L}}

\newcommand{\MeV}{{\rm ~MeV}}
\newcommand{\GeV}{{\rm ~GeV}}
\newcommand{\TeV}{{\rm ~TeV}}

\begin{document}
\leftline{}
\rightline{CP3-20-39, MCnet-20-13, IFJPAN-IV-2021-1}

\title{A quantitative study on helicity inversion in Majorana neutrino decays at the LHC}

\author{Richard Ruiz}
\email{richard.ruiz@uclouvain.be}
\affiliation{Centre for Cosmology, Particle Physics and Phenomenology {\rm (CP3)},\\
Universit\'e Catholique de Louvain, Chemin du Cyclotron, Louvain la Neuve, B-1348, Belgium}
\affiliation{Institute of Nuclear Physics, Polish Academy of Sciences, ul. Radzikowskiego, Cracow 31-342, Poland}

\begin{abstract}
We report an analytical and numerical investigation into the impact of helicity inversion in LHC processes that do not conserve lepton number $(L)$. As a case study, we focus on  the production and decay of Majorana neutrinos $(N)$ through on- and off-shell $W$ bosons in the  Phenomenological  Type I Seesaw model. Using the Monte Carlo event generator \texttt{MadGraph5\_aMC@NLO} in conjunction with the \texttt{HeavyN} model libraries, we perform exact matrix element (ME) computations without the narrow width approximation. Despite helicity inversion appearing explicitly in  MEs, we report the absence of helicity suppression of $L$-violating collider observables for $1\to4$ and $2\to4$ processes that are dominated by resonant $N$ production. We attribute this incongruity to the different scalings of 4-momenta and squared 4-momenta in MEs and squared MEs, with exact cancelations occurring in the latter when $N$ goes on-shell in the small-width limit. In off-shell regimes, total suppression / enhancement of $L$ violation can emerge. Implications for other neutrino mass models are discussed.
\end{abstract}

\date{\today} 

\maketitle

\section{Introduction}\label{sec:intro}

Among the outstanding questions in particle physics~\cite{EuropeanStrategyGroup:2020pow} is whether the light neutrinos observed in nature $(\nu)$ are their own antiparticle, i.e., are they Majorana fermions? If so, then the Lagrangian of the Standard Model  of particle physics  (SM),  which stipulates that neutrinos are massless, must be extended by dimensionful operators that violate the SM's conservation of lepton number $(L)$. Gauge invariance and renormalizability, however, require that such operators have ultraviolet completions, and thereby suggests the possibility of new particles~\cite{Ma:1998dn}. Hence, discovering the  Majorana nature of neutrinos may be a stepping-stone to realizing a mechanism for neutrino mass-generation \cite{Minkowski:1977sc,Konetschny:1977bn,Yanagida:1979as,Glashow:1979nm,Mohapatra:1979ia,GellMann:1980vs,Shrock:1980ct,Schechter:1980gr,Cheng:1980qt,Lazarides:1980nt,Mohapatra:1980yp,Zee:1980ai,Hall:1983id,Foot:1988aq}, new gauge forces \cite{Pati:1974yy,Mohapatra:1974hk,Mohapatra:1974gc,Senjanovic:1975rk,Mohapatra:1979ia,Mohapatra:1980yp}, or even grand unification \cite{Yanagida:1979as,GellMann:1980vs,Lazarides:1980nt,Langacker:1980js,Hall:1983id,Bajc:2006ia}.

Despite this importance, however, 
direct tests of  neutrinos'  Majorana nature, such as through searches for neutrinos' magnetic dipole moments or through $\vert\Delta  L\vert = 2$ transitions like  neutrinoless $\beta\beta$ decay  $(0\nu\beta\beta)$, are encumbered by manifestations of the so-called Dirac-Majorana Confusion Theorem~\cite{Kayser:1982br,Mohapatra:1998rq}.
In the absence of new particles, the theorem in  its standard formulation~\cite{Mohapatra:1998rq} shows that an inherent helicity  inversion in  such processes leads to matrix elements (ME) being proportional to  light  neutrino  masses. This implies  that transition probabilities formally vanish in the limit of vanishing neutrino masses, and thus are classified as being helicity suppressed.  As such, two complementary approaches to the ``Majorana'' question are embraced: The first is the development of large-scale facilities that, for example, aim to  measure the $0\nu\beta\beta$ decay rate. The second relies on direct searches for $\vert\Delta  L\vert = 2$ processes in the context of neutrino mass models.
For reviews of these approaches, see Refs.~\cite{Atre:2009rg,Tello:2010am,Deppisch:2015qwa,Cai:2017mow,Cirigliano:2018yza,Dolinski:2019nrj}.

In the second approach,  processes that do not conserve $L$ are mediated by  new particles~\cite{Minkowski:1977sc,Konetschny:1977bn,Yanagida:1979as,Glashow:1979nm,Mohapatra:1979ia,GellMann:1980vs,Shrock:1980ct,Schechter:1980gr,Cheng:1980qt,Lazarides:1980nt,Mohapatra:1980yp,Zee:1980ai,Hall:1983id,Foot:1988aq} that are  typically much heavier than light neutrinos, but possibly  lighter than the electroweak (EW)  scale. Crucially,  the Confusion Theorem follows from rather generic kinematical arguments in the context of chiral gauge theories, e.g., the EW theory, and not on flavor symmetries  as considered, for example, in Refs.~\cite{Pilaftsis:1991ug,Kersten:2007vk,Antusch:2015mia,Moffat:2017feq}. As such, in scenarios with heavy Majorana neutrinos $(N)$, helicity inversion manifests as asymmetries in angular distributions that distinguish  $\vert\Delta  L\vert = 0$ and $\vert\Delta  L\vert = 2$ channels \cite{Han:2012vk,Chen:2013foz,Gluza:2016qqv,Ruiz:2017nip,Arbelaez:2017zqq,Balantekin:2018azf,Balantekin:2018ukw,Hernandez:2018cgc,Fukuyama:2019jiq}. 
However, while generalizations of the theorem show~\cite{Han:2012vk} that these MEs  are consistently proportional to heavy neutrino  masses  $(m_N)$, past studies have not specifically investigated whether the MEs also vanish when $m_N$ do. For resonantly produced Majorana neutrinos this is pertinent as the often-quoted equality of  $\vert\Delta  L\vert = 0$ and $\vert\Delta  L\vert = 2$ decay rates,  which implies the absence of helicity-suppression, assumes the narrow width approximation and that resonant $N$ can be treated as unpolarized states~\cite{Keung:1983uu,Ferrari:2000sp}. This is despite the presence of chiral couplings and that needed criteria may not be satisfied for currents with Majorana fermions~\cite{Berdine:2007uv,Kauer:2007zc,Kauer:2007nt,Uhlemann:2008pm,Artoisenet:2012st}.

\begin{figure*}
\subfigure[]{\includegraphics[width=0.49\textwidth]{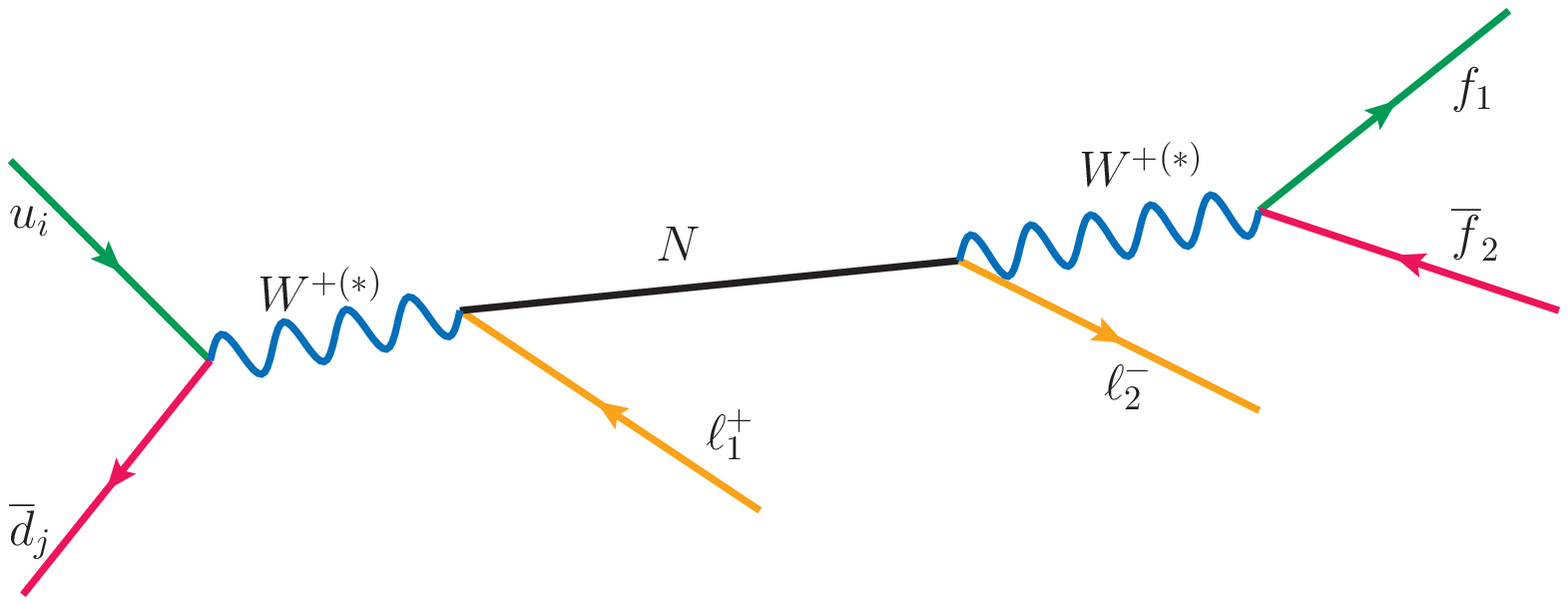}		\label{fig:diagram_HeavyN_DYX_llff_LNC}}
\subfigure[]{\includegraphics[width=0.49\textwidth]{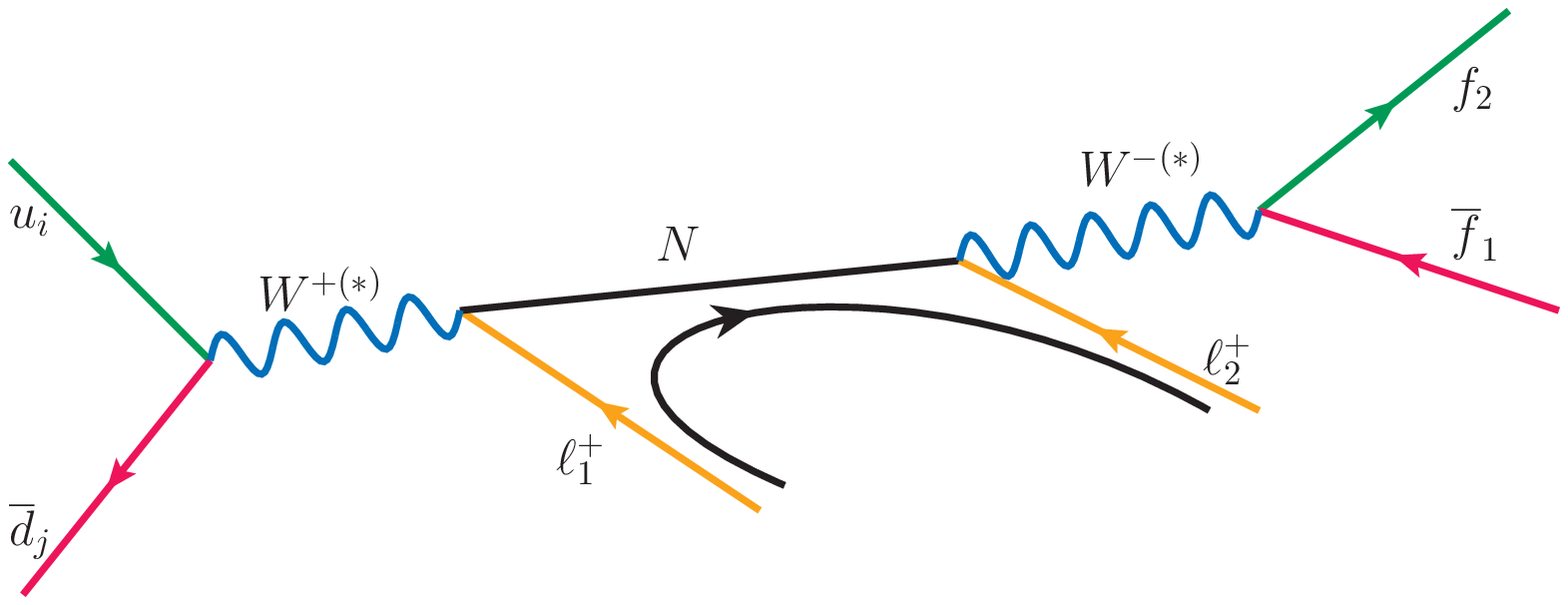}		\label{fig:diagram_HeavyN_DYX_llff_LNV}}
\caption{
 Born-level, diagrammatic representation of the (a) $L$-conserving process  $u\overline{d} \to W^{+} \to N \ell_1^+ \to  \ell_1^+ \ell_2^- f_1 \overline{f_2}$, and (b) its $L$-violating analogue $u\overline{d} \to W^{+} \to N \ell_1^+ \to  \ell_1^+ \ell_2^+ \overline{f_1} f_2$. Interfering diagrams not shown. Drawn with \texttt{JaxoDraw}~\cite{Binosi:2008ig}.
}
\label{fig:diagram_HeavyN_DYX_llff}
\end{figure*}

In this study, we report an analytical and numerical investigation into the impact of helicity inversion  in $L$-violating transition rates involving heavy Majorana neutrinos at the $\sqrt{s}=13\TeV$ Large Hadron Collider (LHC). As a representative case study, we work in the framework of the Phenomenological Type I Seesaw model and focus on the $L$-violating decay and scattering processes~\cite{Keung:1983uu}
\begin{align}
	W^\pm \quad  &\to e_1^\pm N^{(*)} \to e_1^\pm e_2^\pm j j,	\label{eq:main_decay} \\
p p  \to W^{\pm (*)} &\to e_1^\pm N^{(*)} \to e_1^\pm e_2^\pm j j,	\label{eq:main_scatter}
\end{align}
and their $L$-conserving counterparts, as shown diagrammatically at the parton level in Fig.~\ref{fig:diagram_HeavyN_DYX_llff}. While Eqs. \eqref{eq:main_decay} and \eqref{eq:main_scatter} are intimately related, their individual considerations explore subtle polarization and virtuality effects.

By performing exact ME computations with the Monte Carlo (MC) event generator  \texttt{MadGraph5\_aMC@NLO}~\cite{Stelzer:1994ta,Alwall:2014hca}
in conjunction with the \texttt{HeavyN} model libraries~\cite{Alva:2014gxa,Degrande:2016aje},
 and without invoking the narrow width approximation, we find that the helicity suppression in collider observables
 is numerically negligible for processes driven by resonant production of Majorana neutrinos 
 with masses $m_N > 1\GeV$ and total widths $\Gamma_N \ll m_N$. We attribute the seeming incongruity with the presence of helicity inversion  to the different scaling of 4-momenta and squared 4-momenta in MEs and squared MEs. In the on-shell, small-width limit, this leads to cancelations of the dependence on $m_N$, with corrections proportional to  off-shell virtualities and $\Gamma_N$. Outside this limit we observe the opposite behavior. When the off-shell behavior is driven by a large width, we find that the $\vert\Delta  L\vert = 2$  channel is helicity-suppressed; when the off-shell behavior is driven by a too large mass, helicity enhancement emerges.  As the arguments here are kinematical in nature,  analogous findings apply to other models with Majorana $N$.

This study  continues in the following order:
In Sec.~\ref{sec:theory_model}  we summarize the theoretical framework in which we work.
In Sec.~\ref{sec:setup} we document our  computational setup.
We then identify analytically in Sec.~\ref{sec:helicityTheory} the helicity inversion at the ME level, its propagation to the squared ME level, and its (mis)cancelation in the (off)on-shell limit for the processes in Fig.~\ref{fig:diagram_HeavyN_DYX_llff}.
We also comment on implications for other models and $\vert \Delta L\vert = 2$  processes.
We present our numerical comparisons in Sec.~\ref{sec:results}
and  conclude  in Sec.~\ref{sec:conclude}.

\section{Theoretical Framework}\label{sec:theory_model}
To investigate the potential helicity suppression of  $L$-violating processes mediated by Majorana neutrinos,  we work in the framework of the Phenomenological Type I Seesaw. In this well-documented ~\cite{delAguila:2008cj,Atre:2009rg,Pascoli:2018heg} scenario, the masses and mixing angles of light  $(\nu_k)$ and heavy $(N_{k'})$ neutrino mass eigenstates are decoupled in order to conduct flavor model-independent studies and searches.

In  this model, the SM's field content is extended by $n_R\geq3$ right-handed (RH) neutrinos $(\nu_R^i)$ that are gauge-singlets, i.e., are chargeless / sterile, under the SM gauge interactions. This allows the $\nu_R^i$ to possess RH Majorana masses $(\mu_R^{ij})$, which violate $L$ conservation and can, in principle, can acquire any value\footnote{If coupled to other physics, e.g., particle dark matter~\cite{Asaka:2005pn,Asaka:2005an} or global symmetries~\cite{Mohapatra:1986aw,Mohapatra:1986bd,Bernabeu:1987gr,Pilaftsis:1991ug,Kersten:2007vk,Antusch:2015mia,Moffat:2017feq}, then the values of $\mu_R^{ij}$ can be stringently constrained.}.
The decoupling of $\mu_R^{ij}$ subsequently suppresses light neutrino masses~\cite{Minkowski:1977sc,Yanagida:1979as,GellMann:1980vs,Glashow:1979nm,Mohapatra:1979ia,Shrock:1980ct,Schechter:1980gr} and is distinct from other neutrino mass mechanisms, e.g., the Type II Seesaw~\cite{Konetschny:1977bn,Schechter:1980gr,Cheng:1980qt,Lazarides:1980nt}, where $\nu$ masses are generated via left-handed (LH) Majorana masses.

Accordingly, the Lagrangian of the Phenomenological Type I Seesaw $(\mathcal{L}_{\rm Type~I})$ is characterized  by extending the SM  Lagrangian $(\mathcal{L}_{\rm SM})$ at the renormalizable level by  kinetic and mass  terms for the $\nu_R^i$ $(\mathcal{L}_{\rm Kin.})$, and by Yukawa couplings $(\mathcal{L}_{\rm Y})$ between the $\nu_R^i$, the SM Higgs field  $(\Phi)$, and the SM's LH lepton doublets $L^{jT} = (\nu_L^{j}, l_L^j)$,
\begin{equation}
\mathcal{L}_{\rm Type~I} = \mathcal{L}_{\rm SM} + \mathcal{L}_{\rm Kin.} +\mathcal{L}_{\rm Y}.
\end{equation}

After EWSB and diagonalizing charged  lepton  flavor states into their  mass eigenstates $(\ell = e,\mu,\tau)$, the flavor eigenstates of active,  LH neutrinos $(\nu_{L\ell})$ can be decomposed into mass  eigenstates via the rotation~\cite{Atre:2009rg}
\begin{equation}
\nu_{L \ell} = \sum_{k=1}^3 U_{\ell k} \nu_k + \sum_{k'=1}^{n_R} V_{\ell k'} N_{k'}.
\label{eq:nuMix}
\end{equation}
Here the complex-valued mixing elements $U_{\ell k}$ and $V_{\ell k'}$ parametrize the mixing between the flavor state $\nu_{L\ell}$   with the mass eigenstates $\nu_k$ and $N_{k'}$.
For updated measurements and constraints of mixing angles, see Refs.~\cite{Blennow:2016jkn,Fernandez-Martinez:2016lgt,Dentler:2018sju,Esteban:2020cvm}.

Given Eq.~\eqref{eq:nuMix},  the relevant interaction Lagrangian describing the charged current interactions of $N_{k'}$ is,
\begin{eqnarray}
\mathcal{L} &=& -\frac{g_W}{\sqrt{2}}W^+_\mu 
\sum_{\ell}^\tau\left[\overline{\nu_{L\ell}}\gamma^\mu P_L \ell\right] +\text{H.c.}\quad
\\
&=& -\frac{g_W}{\sqrt{2}}W^+_\mu 
\sum_{k=1}^3
\sum_{\ell}^\tau\left[\overline{\nu_k}U^*_{\ell k}\gamma^\mu P_L \ell\right]
\nonumber\\
& & 
 -\frac{g_W}{\sqrt{2}}W^+_\mu 
\sum_{k'=1}^{n_R}
\sum_{\ell}^\tau\left[\overline{N_{k'}}V^*_{\ell k'}\gamma^\mu P_L \ell\right]
+\text{H.c.}\quad
\label{eq:nuLag}
\end{eqnarray}
Here, $g_W=e/\sin\theta_W\approx0.65$ is the usual weak gauge coupling constant in the SM, 
and $P_{L/R}=(1/2)(1\mp\gamma^5)$ are the LH/RH chiral projection operators in four-component notation. Using Eq.~\eqref{eq:nuMix} to make analogous substitutions, interaction Lagrangians involving the $Z$  and Higgs can be built accordingly~\cite{delAguila:2008cj,Atre:2009rg}.
Throughout this study we consider the impact of only  the lightest heavy  mass eigenstate $(N_{k'=1})$,
which we relabel as $N\equiv N_{k'=1}$ with $V_{\ell N} \equiv V_{\ell k'=1}$.
We do so to isolate the impact of helicity inversion in $L$-violating currents that can otherwise be obfuscated by strong interference.

\section{Computational Setup}\label{sec:setup}
We now briefly document the computational setup of this study. After summarizing the MC setup in Sec.~\ref{sec:setup_mc}, the numerical inputs for SM and heavy neutrino parameters are respectively provided in Secs.~\ref{sec:setup_sm} and \ref{sec:setup_nx}.

\subsection{Monte Carlo Setup}\label{sec:setup_mc}
To perform our numerical computations, we use the MC event generator \texttt{MadGraph5\_aMC@NLO} (v2.7.0) \cite{Stelzer:1994ta,Alwall:2014hca} (\texttt{mgamc}). The simulation suite~\cite{Stelzer:1994ta,Maltoni:2002qb,deAquino:2011ub,Artoisenet:2012st,Hirschi:2011pa,Alwall:2014bza,Alwall:2014hca} operates by constructing helicity amplitudes for short-distance decay and scattering processes~\cite{Murayama:1992gi,Stelzer:1994ta,deAquino:2011ub} according formalism of Refs.~\cite{Hagiwara:1985yu,Hagiwara:1986vm,Hagiwara:1988pp,Murayama:1992gi}  and performs fast numerical integration over phase space through  MC sampling~\cite{Maltoni:2002qb,Alwall:2011uj}.
For heavy neutrino interactions governed by  the Lagrangian of Eq.~\eqref{eq:nuLag}, we import into \texttt{mgamc} the \texttt{HeavyN}~\cite{Alva:2014gxa,Degrande:2016aje} \texttt{FeynRules}  ~\cite{Christensen:2008py,Degrande:2011ua,Alloul:2013bka}  libraries. This employs the conventions   for Majorana currents developed in  Refs.~\cite{Denner:1992vza,Denner:1992me}.
For select calculations, we compute helicity-polarized MEs in \texttt{mgamc} according to the formalism of Ref.~\cite{BuarqueFranzosi:2019boy}.

\subsection{Standard Model  Inputs}\label{sec:setup_sm}
For numerical computations we work in the  $n_f=5$  massless / active quarks scheme with SM inputs set to
\begin{align}
m_t(m_t) = 173.3\GeV, &  ~ M_Z = 91.1876\GeV, \\
\alpha^{-1}_{\rm QED}(M_Z) = 127.94, &  ~ G_F = 1.174560\times10^{-5}\GeV^{-2}.
\end{align}
We take the  Cabbibo-Kobayashi-Maskawa matrix equal to the identity matrix.
For relevant computations we use the MSTW 2008  leading order parton density functions (\texttt{lhaid=21000}) \cite{Martin:2009iq} as evolved by \texttt{LHAPDF} (v6.2.3)~\cite{Buckley:2014ana},
and set the collinear factorization scale $(\mu_f)$ to
\begin{equation}
 \mu_f = M_W\approx 79.95\GeV.
 \end{equation}

\subsection{Heavy Neutrino Inputs}\label{sec:setup_nx}

In addition to SM inputs, the  (default) \texttt{HeavyN} model libraries \cite{Alva:2014gxa,Degrande:2016aje} consists of three Majorana neutrino mass eigenstates $N_{k'}$ with mass eigenvalues   $m_{N_{k'}}$ and active-sterile mixing elements $V_{\ell k'}$ associated with lepton flavor $\ell$.  As explained at the end of Sec.~\ref{sec:theory_model}, we decouple two  $N_k$  in order to  isolate helicity inversion in the absence of interference.  To do this  numerically, we set
\begin{equation}
m_{N_2}, m_{N_3} =  10^{12}\GeV ~\text{and}~  \vert V_{\ell  2}\vert, \vert V_{\ell  3}\vert = 10^{-10}.
\end{equation}
As the  values of $m_{N_1}, \vert V_{\ell  N_1}\vert$ are varied, the total width $(\Gamma_{N_1})$ of $N_1$ is evaluated\footnote{
We note that there is a limited ability in \texttt{MadDecay} to compute extremely small particle widths, which can occur for particularly tiny mixing elements. In this study, no such threshold was reached. However, a possible workaround for future studies would be to evaluate a total width at an artificially large mixing element and rescale to much smaller ones. For example: one can compute $\Gamma_N(\vert V_{\ell N}\vert=10^{-3}) = (10^{-3}/10)^2 \times \Gamma_N(\vert V_{\ell N}\vert=10)$.} on-the-fly using \texttt{MadDecay}~\cite{Alwall:2014bza}.

\section{Helicity inversion in matrix elements with Majorana Neutrinos} \label{sec:helicityTheory}
  
For $\vert\Delta L\vert =2$ transitions that are mediated by heavy Majorana neutrinos in the Phenomenological Type I Seesaw, we establish  in this  section  the presence of helicity inversion in MEs and its propagation into squared MEs. While the inversion has far-reaching consequences \cite{Kayser:1982br,Mohapatra:1998rq,Han:2012vk,Chen:2013foz,Gluza:2016qqv,Ruiz:2017nip,Arbelaez:2017zqq,Balantekin:2018azf,Balantekin:2018ukw,Hernandez:2018cgc,Fukuyama:2019jiq},
it is essentially  a quirk of chiral gauge theories, like the EW theory, and follows from the Charge-Parity-Time (CPT) theorem. We organize this derivation by first considering $L$-conserving, $4$-body decays of $W$ bosons in Secs.~\ref{sec:helicity_lnc}, and then $L$-violating decays in Secs.~\ref{sec:helicity_lnv}.  We draw special attention to the precise origin of the inversion and the scaling of (squared) momenta in (squared) MEs. In Sec.~\ref{sec:helicity_scatt} we consider analogous processes  in $2\to4$ scattering and comment on the implications for other neutrino mass models in Sec.~\ref{sec:helicity_other}.

\subsection{$W$ boson decays with $\Delta L = 0$}\label{sec:helicity_lnc}

As a first step to studying helicity inversion in $\vert \Delta L\vert = 2$ processes, we consider  the following $L$-conserving, $4$-body $W$ boson decay mediated by a Majorana neutrino $N$,
\begin{align}
W^+_{\lambda_W}(p_W) 	\to & ~\ell_{R1}^+(p_1) ~N_{\lambda_N}(p_N)  \\
					\to & ~\ell_{R1}^+(p_1) ~\ell_{L2}^-(p_2) ~c_L(p_c) ~\overline{s_R}(p_s),
					\label{eq:proc_wdecay_chain_lnc}
\end{align}
as shown as a sub-process in Fig.~\ref{fig:diagram_HeavyN_DYX_llff_LNC}. 
Here, the subscripts $\lambda_W=0,\pm1$ and $\lambda_N=L,R$ denote the helicities of $W^+$ and $N$. The helicities of  massless fermions $\ell_k,  c, s$ are fixed by the $W$  boson's chiral couplings.

Working in the unitary gauge and in the \texttt{HELAS} basis~\cite{Murayama:1992gi} for helicity amplitudes, the corresponding  ME is 
\begin{align}
-i\mathcal{M}_\lnc^W = & ~ \varepsilon_\mu(p_W,\lambda_W) 
~T_\lnc^{\nu\mu}(p_1,p_2,p_N) \nonumber\\
& ~ \times  \Delta_{\nu \rho}(p_c+p_s) ~J^\rho(p_c,p_s).
\label{eq:helicityAmp_lnc_full}
\end{align}
Here the $(c\overline{s})$  fermion current and $W$ propagator are
\begin{align}
J^\rho(p_c,p_s) &= \frac{-i g_W \delta^{AB}}{\sqrt{2}} \left[\overline{u}^A_L(p_c)\gamma^\rho P_L v^B_R(p_s)\right],
\\
 \Delta_{\nu \rho}(k)  &= \cfrac{-i(g_{\nu\rho}-k_\nu k_\rho / M_W^2)}{(k^2 - M_W^2 + i\Gamma_W M_W)},
 \label{eq:helicityAmp_WProp}
\end{align}
and the $L$-conserving $(\ell_1^+  N \ell_2^-)$ lepton current  is
\begin{align}
 T_\lnc^{\nu\mu}&(p_1,p_2,p_N)  = \nonumber\\
&  \qquad ~ \qquad \left(\frac{-i g_W}{\sqrt{2}}\right)^2  V_{\ell_1 N}^* V_{\ell_2 N} 
\times 
\mathcal{N}^{\nu\mu}_\lnc \times  \mathcal{D}, 
\\
 \mathcal{N}^{\nu\mu}_\lnc &= \left[\overline{u}_L(p_2)\gamma^\nu P_L(\not\!p_N + m_N\mathbb{I}_4)\gamma^\mu P_L v_R(p_1) \right ], \label{eq:helicityAmp_lnc}  
\\
 \mathcal{D} &= 
\frac{i}{(p_N^2 - m_N^2 + i \Gamma_N m_N)}.
\label{eq;pole}
\end{align}
In the quark current $J^\rho$, the indices $A,B=1,\cdots, N_c=3$ run over the QCD color states, and the Kronecker $\delta$-function  $\delta^{AB}$ ensures a color-singlet $W^*\to q\overline{q'}$ splitting. In the lepton current $T_\lnc^{\nu\mu}$, 
$\mathcal{D}$ is the pole structure of the Breit-Wigner propagator for the Majorana neutrino $N$.

Importantly, the $(\ell_1^+  N \ell_2^-)$ fermion current is initiated/terminated by successive $W$  interactions. These are maximally parity-violating, are oriented in the left chiral direction, and  are responsible for the two $(\gamma^\alpha  P_L)$ in  $\mathcal{N}^{\nu\mu}_\lnc$. Due to orthogonality of RH and LH chiral projection operators, the intermediate $N$ is confined  to its LH helicity state (the $\not\!p_N$ term). The transition is helicity conserving as  RH helicity states  (the $m_N\mathbb{I}_4$ term) do not contribute to successive chiral interactions with the same chirality.

After anticommuting the left-most $P_L$ and using na\"ive power counting to extract the energy  dependence  from  spinors, we obtain for the $(\ell_1^+  N \ell_2^-)$ lepton current:
\begin{align}
\mathcal{N}^{\nu\mu}_\lnc =& \left[\overline{u}_L(p_2)\gamma^\nu P_L(\not\!p_N + m_N\mathbb{I}_4)\gamma^\mu P_L v_R(p_1) \right ]
\\
=&  
\left[\overline{u}_L(p_2)\gamma^\nu  \not\!p_N \gamma^\mu P_L v_R(p_1) \right ]  \\
\sim& ~ 
 \sqrt{E_2} ~ E_N ~ \sqrt{E_1} \sim M_W^2.
 \label{eq:helicityAmp_lnc_reduced}
\end{align}
The scaling in the last line  shows that the amplitude $\mathcal{M}_\lnc^W$ for the $W^+ \to  \ell_1^+ \ell_2^- c\overline{s}$  decay grows  with the  energy of $N$, and therefore is  {not}  suppressed for vanishing $m_N$. 

We note that due to Lorentz invariance the scaling of 4-momenta $(p^\mu)$ and their squares $(p^\mu  p_\mu)$ differ.
Importantly, this  leads to different behavior in squared MEs than in Eq.~\eqref{eq:helicityAmp_lnc_reduced}. 
In particular, one finds using Ref.~\cite{Han:2012vk},
\begin{align}
\sum \vert \mathcal{M}_\lnc^W \vert^2 & \sim 
\sum T_\lnc^{\nu\mu} \left[T_\lnc^{\alpha\beta}\right]^\dagger
\\
& \sim  \sum \mathcal{N}^{\nu\mu}_\lnc \left[\mathcal{N}^{\alpha\beta}_\lnc\right]^\dagger
\times \vert \mathcal{D}(p_N^2) \vert ^2, 
  \end{align}
  where the squared and spin-summed current scales as
  \begin{align}
\sum \mathcal{N}^{\nu\mu}_\lnc \left[\mathcal{N}^{\alpha\beta}_\lnc\right]^\dagger
 \sim   E_2 ~ p_N^2 ~ E_1  \sim ~ M_W^2 p_N^2.
 \label{eq:helicity_ME2_inc_scale}
  \end{align}
  
  Interestingly, Eq.~\eqref{eq:helicity_ME2_inc_scale} shows that the squared ME scales as the virtuality of the intermediate $N$, and can potentially vanish for tiny $m_N$ in the on-shell limit. 
In this region of phase space however, i.e.,  when
  \begin{equation}
  \delta p_N^2 \equiv \vert p_N^2 - m_N^2 \vert \ll  \Gamma_N m_N \ll m_N^2,
  \label{eq:massNeighborhood}
  \end{equation}  
  the pole structure of the propagator $\mathcal{D}$ behaves as
  \begin{align}
\vert  \mathcal{D}(p_N^2) \vert ^2 &= \cfrac{1}{(p_N^2 - m_N^2)^2 + (\Gamma_N m_N)^2}
  \\ 
  & = \cfrac{1}{(\Gamma_N m_N)^2\left[ 1 + \frac{\delta p_N^4}{(\Gamma_N m_N)^2} \right]}
  \\
  &= \cfrac{1}{(\Gamma_N m_N)^2}\left[ 1 - \mathcal{O}\left(\frac{\delta p_N^4}{(\Gamma_N m_N)^2}\right) \right].
  \end{align}
  In combination with the scaling in Eq.~\eqref{eq:helicity_ME2_inc_scale}, 
  we obtain\footnote{
  We note that the dependence of $\sum \vert \mathcal{M}_\lnc^W \vert^2 $ here on $M_W^2$  does not account for contributions from  $\varepsilon_\mu$, $\Delta_{\nu\rho}$, and $J^\rho$ in Eq.~\eqref{eq:helicityAmp_lnc_full}. Throughout this entire section we suppress these extra factors.}
  \begin{align}
\sum \vert \mathcal{M}_\lnc^W \vert^2 & \sim 
 \cfrac{M_W^2 p_N^2}{(\Gamma_N m_N)^2}\left[ 1 - \mathcal{O}\left(\frac{\delta p_N^4}{(\Gamma_N m_N)^2}\right) \right],
 \label{eq:meSq_lnc_leading}
  \end{align}
and see  that the  dependence on $m_N^2$ is actually cancelled in the on-shell limit. Hence, like at the ME level, the leading contribution to the $W^+ \to  \ell_1^+ \ell_2^- c\overline{s}$  decay rate at the squared ME level does not vanish for vanishing $m_N$.

\subsection{$W$ boson decays with $\vert \Delta L\vert = 2$}\label{sec:helicity_lnv}

Moving to the $L$-violating analogue of the process in Eq.~\eqref{eq:proc_wdecay_chain_lnc}, we have the $4$-body $W$ boson decay chain
\begin{align}
W^+_{\lambda_W}(p_W) 	\to & ~\ell_{R1}^+(p_1) ~N_{\lambda_N}(p_N)  \\
					\to & ~\ell_{R1}^+(p_1) ~\ell_{R2}^+(p_2) ~\overline{c}_R(p_c) ~{s_L}(p_s),
\label{eq:proc_wdecay_chain_lnv}					
\end{align}
as shown as a sub-process in Fig.~\ref{fig:diagram_HeavyN_DYX_llff_LNV}. 
Following the same procedure as needed to construct $\mathcal{M}_\lnc^W$ in Eq.~\eqref{eq:helicityAmp_lnc_full}, the ME of the $L$-violating decay process is given by
\begin{align}
-i\mathcal{M}_\lnv^W = ~ & \varepsilon_\mu(p_W,\lambda_W) 
~T_\lnv^{\nu\mu}(p_1,p_2,p_1+p_c+p_s) \nonumber\\
& \times  \Delta_{\nu \rho}(p_c+p_s) ~J^\rho(p_s,p_c) \nonumber\\
& + (p_1\leftrightarrow p_2).
\label{eq:helicityAmp_lnv_full}
\end{align}
Up to external momentum reassignments, the quark current $J^\rho$, polarization vector $\varepsilon_\mu$, and propagator $\Delta_{\nu\rho}$ are the same as in the $L$-conserving case.
In the last line is the interference from $\ell_1\leftrightarrow\ell_2$ particle exchange. Due to charge  conservation, no second term exists in $\mathcal{M}_\lnc^W$. 

The key difference from the $L$-conserving ME is the $L$-violating $(\ell_1^+  N \ell_2^+)$ fermion current. To derive this we note that in going from the $W^+ \to  \ell_1^+ \ell_2^- c\overline{s}$ process to $W^+ \to  \ell_1^+ \ell_2^+ \overline{c} s$, one effectively imposes a charge inversion on the electrically neutral $(\ell_2^- c \overline{s})$ system.
Under CPT, this is the same as a parity-time inversion and, significantly, is  expressible as Feynman rules~\cite{Denner:1992vza,Denner:1992me}.

These state that after assuming a  \textit{fermion flow} (curve in Fig.~\ref{fig:diagram_HeavyN_DYX_llff_LNV}) the  $(N\ell_2^+W^-)$ vertex as derived from the Lagrangian in Eq.~\eqref{eq:nuLag} is parity-inverted and becomes
\begin{equation}
-\frac{ig_W}{\sqrt{2}} V_{\ell_2 N} \gamma^\nu P_L ~\to~ 
(-1)^2\frac{ig_W}{\sqrt{2}}V_{\ell_2 N} \gamma^\nu P_R.
\label{eq:feynmanrule_01}
\end{equation}
Consistently, as $\ell_2$'s own fermion number flow, which points inward, is antiparallel to the conventional fermion flow, which points outward, its spinor is time-inverted: 
\begin{equation}
\overline{v}_R(p_2) ~\to~ \overline{u}_R(p_2). 
\label{eq:feynmanrule_02}
\end{equation}

Propagating these modifications and defining for compactness $\tilde{p}_k \equiv p_k+p_c+p_s$,
for $k=1,2$, we find that both $L$-violating  $(\ell_1^+  N \ell_2^+)$ fermion currents are given by
\begin{align}
 T_\lnv^{\nu\mu}&(p_1,p_2,\tilde{p}_k) = \nonumber\\
&    -\left(\frac{-i g_W}{\sqrt{2}}\right)^2  V_{\ell_1 N}^* V_{\ell_2 N} 
\times 
\mathcal{N}^{\nu\mu}_\lnv \times  \mathcal{D}(\tilde{p}_k^2), 
\\
 \mathcal{N}^{\nu\mu}_\lnv &= \left[\overline{u}_R(p_2)\gamma^\nu P_R(\not\!\tilde{p}_k + m_N\mathbb{I}_4)\gamma^\mu P_L v_R(p_1) \right ]. 
 \label{eq:helicityAmp_lnv}  
\end{align}
Importantly, this differs from the $L$-conserving analogue  $ \mathcal{N}^{\nu\mu}_\lnc$ in Eq.~\eqref{eq:helicityAmp_lnc} by the replacement of the leftmost chiral projection operator $P_L$ with the RH projector $P_R$, a consequence of Eq.~\eqref{eq:feynmanrule_01}.
Using again the orthogonality of projection operators we see that the intermediate $N$ is confined  to its RH helicity state (the $m_N\mathbb{I}_4$ term).
The $L$-violating transition is helicity inverting as LH helicity states  (the $\not\!p_N$ term) do not contribute to successive chiral interactions of opposite chirality.

After anticommuting the operator $P_R$, we obtain  
\begin{align}
 \mathcal{N}^{\nu\mu}_\lnv =& ~ \left[\overline{u}_R(p_2)\gamma^\nu P_R(\not\!\tilde{p}_k + m_N\mathbb{I}_4)\gamma^\mu P_L v_R(p_1) \right ] \\
=&  ~ 
m_N \times \left[\overline{u}_L(p_2)\gamma^\nu   \gamma^\mu P_L v_R(p_1) \right ]  \\
\sim& ~ m_N  \sqrt{E_2}  \sqrt{E_1} \sim m_N M_W.
 \label{eq:helicityAmp_lnv_reduced}
\end{align}
In the last line we again employ  na\"ive power counting to find that both $(\ell_1^+  N \ell_2^+)$ currents are proportional to $m_N$, independent of $\tilde{p}_k$.  Subsequently, we see that both currents vanish for vanishing Majorana neutrino mass, in line with expectations from the Confusion Theorem.

To address the pole structure in the Majorana neutrino's propagator ($\mathcal{D}$ in Eq.~\eqref{eq;pole}) as we did for the $L$-conserving decay, we consider again when $N$ is  (nearly) on-shell.  Without the loss of generality, we assume $\tilde{p}_1^2 = (p_1 + p_c+p_s)^2$ satisfies the near on-shell condition of Eq.~\eqref{eq:massNeighborhood}. By momentum conservation, the non-resonant momentum configuration  has the virtuality
\begin{equation}
\tilde{p}_2^2 = (p_W - p_1)^2 = M_W^2 - 2 M_W E_1. 
\end{equation}
For these configurations of $\tilde{p}_k$, we obtain the expansions
  \begin{align}
  \mathcal{D}(\tilde{p}_1)  &= \cfrac{i}{(\tilde{p}_1^2 - m_N^2) + i(\Gamma_N m_N)}
  \\
  &= \cfrac{1}{\Gamma_N m_N}\left[ 1 - \mathcal{O}\left(\frac{\delta p_N^2}{\Gamma_N m_N}\right) \right],
  \end{align}
  \begin{align}
  \mathcal{D}(\tilde{p}_2)  &= \cfrac{i}{(\tilde{p}_2^2 - m_N^2) + i(\Gamma_N m_N)}  
  \\
  &= \cfrac{i}{M_W^2\left(1 - 2\frac{E_1}{M_W}  - \frac{m_N^2 - i(\Gamma_N m_N)}{M_W^2}\right)}
  \\
    &= \cfrac{i}{M_W^2}\left[ 1 + \mathcal{O}\left(\frac{E_1}{M_W}\right) + \mathcal{O}\left(\frac{m_N^2}{M_W^2}\right) \right].
  \end{align}
After combining $\mathcal{D}(\tilde{p}_k)$ with $\mathcal{N}^{\nu\mu}_\lnv$, we see that the $m_N$ dependence in the non-resonant contribution scales as
\begin{equation}
 \mathcal{N}^{\nu\mu}_\lnv \times \mathcal{D}(\tilde{p}_2)  \sim 
 i\cfrac{m_N}{M_W}\left[ 1 + \mathcal{O}\left(\frac{E_1}{M_W}, \frac{m_N^2}{M_W^2}\right)
\right],
\end{equation}
and thereby vanishes in the limit that $(m_N/M_W)\rightarrow0$. On the other hand, for the resonant contribution, we obtain a qualitatively different behavior, namely that
\begin{equation}
 \mathcal{N}^{\nu\mu}_\lnv \times \mathcal{D}(\tilde{p}_1)  \sim 
  \cfrac{M_W}{\Gamma_N}\left[ 1 - \mathcal{O}\left(\frac{\delta p_N^2}{\Gamma_N m_N}\right) \right].
  \label{eq:me_lnv_leading}
\end{equation}
This shows that the dependence on $N$'s mass cancels in the resonant contribution and hence generates a non-zero ME for $W^+ \to  \ell_1^+ \ell_2^+ \overline{c}{s}$, even for vanishing $m_N$. While helicity inversion exists at the ME level, its impact is mitigated  by the propagator in the on-shell limit, i.e., when $N$ can be approximated as an asymptotic state. Notably, this is independent of active-sterile mixing.

Moreover, since the ME for  $W^+ \to  \ell_1^+ \ell_2^+ \overline{c}{s}$ scales as the $(\ell_1^+  N \ell_2^+)$ current and its crossing interference, 
\begin{equation}
\mathcal{M}_\lnv^W \sim 
\left[
 \mathcal{N}^{\nu\mu}_\lnv \times \mathcal{D}(\tilde{p}_1) +  \mathcal{N}^{\nu\mu}_\lnv \times \mathcal{D}(\tilde{p}_2)\right],
\end{equation}
we find that the resonant, interference, and non-resonant terms respectively contribute to the squared ME  as 
\begin{equation}
 \vert \mathcal{M}_\lnv^W \vert^2 \sim 
\underset{\rm resonant}{\underbrace{\mathcal{O}\left(\frac{M_W^2}{\Gamma_N^2}\right)}}
+
\underset{\rm interference}{\underbrace{\mathcal{O}\left(\frac{m_N}{\Gamma_N}\right)}}
+
\underset{\rm non-res.}{\underbrace{\mathcal{O}\left(\frac{m_N^2}{M_W^2}\right)}}.
\end{equation}
This tells us that while the non-resonant contribution is negligible compared to the (leading) resonant contribution, the (sub-leading) interference is not guaranteed to be negligible if $m_N \sim M_W$. However, for $m_N\ll M_W$, the total width of $N$ scales as  $\Gamma_N \sim G_F^2 m_N^5 \vert V_{\ell N}\vert^2$, and suggests a numerically insignificant  interference term.

Using Eq.~\eqref{eq:me_lnv_leading} to keep track of formally sub-leading terms, one finds a more exact scaling of the squared ME:
\begin{align}
\sum \vert \mathcal{M}_\lnv^W \vert^2 &\sim 
  \cfrac{M_W^2}{\Gamma_N^2}\left[ 1 - \mathcal{O}\left(\frac{\delta p_N^4}{(\Gamma_N m_N)^2}\right) \right] 
  \nonumber\\
& +
\mathcal{O}\left(\frac{m_N}{\Gamma_N}\right)
+
\mathcal{O}\left(\frac{m_N^2}{M_W^2}\right).
\label{eq:meSq_lnv_leading}
\end{align}
In comparison to the squared ME in Eq.~\eqref{eq:meSq_lnc_leading}, the above demonstrates that in the limit that the Majorana neutrino goes on-shell, the leading contribution to the squared ME of the 
$L$-violating decay  $W^+ \to  \ell_1^+ \ell_2^+ \overline{c}{s}$ admits a dependence on the mass $m_N$ that is identical to that found in the $L$-conserving decay  $W^+ \to  \ell_1^+ \ell_2^- c\overline{s}$. 

Furthermore, for both decay processes, the respective contributions from the polarization vector $\varepsilon_\mu$, the $W^\mp$ propagator $\Delta_{\nu\rho}$, and the $(c\overline{s})/(\overline{c}s)$ current $J^\rho$ in Eqs.~\eqref{eq:helicityAmp_lnc_full} and \eqref{eq:helicityAmp_lnv_full} are the same.
It follows that the squared MEs for the two processes do not just have the same scaling dependence on $m_N$ and $\Gamma_N$ but are, in fact, equal in the limit that $N$ goes on-shell, up to off-shell and finite width corrections.  Therefore, after phase space integration, one can anticipate highly comparable decay rates despite the relative presence of helicity inversion.

\subsection{$W$ boson decays with off-shell $N$}\label{sec:helicity_offshell}

An important qualification for the above result is Eq.~\eqref{eq:massNeighborhood}, which stipulates that the internal Majorana neutrino in the $1\to4$-body decay is or nearly is on its mass shell. Indeed, when comparing the $L$-conserving and $L$-violating squared MEs in Eqs.~\eqref{eq:meSq_lnc_leading} and \eqref{eq:meSq_lnv_leading}, one sees that the  dependence on the neutrino's mass and width only match in this limit. Outside this kinematic limit, mismatches emerge.  While a systematic investigation of off-shell Majorana neutrinos in $\Delta L = 0$ and $\vert \Delta L\vert = 2$ processes is outside the scope of the present work, we can nevertheless outline some generic features.

If $N$ couples to additional new particles, for example to new Higgs or gauge bosons~\cite{Pati:1974yy,Mohapatra:1974hk,Mohapatra:1974gc,Senjanovic:1975rk,Mohapatra:1979ia,Mohapatra:1980yp}, then its width can be much larger than anticipated by the Lagrangian in Sec.~\ref{sec:theory_model}. In particular, if $N$ is light but has a width comparable to its mass, then the ``on-shell'' condition, 
  \begin{equation}
\delta p_N^2 \equiv \vert p_N^2 - m_N^2 \vert \ll  \Gamma_N m_N \sim m_N^2 < M_W^2,
  \end{equation}  
only weakly constraints the virtuality of the internal Majorana neutrino. By taking the difference $(\Delta_\mathcal{M}^{\rm Large~Width})$ of the leading contributions in Eqs.~\eqref{eq:meSq_lnc_leading} and \eqref{eq:meSq_lnv_leading}, 
\begin{align}
&\Delta_\mathcal{M}^{\rm Large~Width}  \equiv \sum \vert \mathcal{M}_\lnc^W \vert^2 - \sum \vert \mathcal{M}_\lnv^W \vert^2 \\
&\sim 
\cfrac{M_W^2}{\Gamma_N^2}
\left( \cfrac{p_N^2}{m_N^2} - 1 \right) \times
 \left[ 1 - \mathcal{O}\left(\frac{\delta p_N^4}{(\Gamma_N m_N)^2}\right) \right],
 \label{eq:asymm_offshell_largeWidth}
\end{align}
a nonzero resultant emerges an grows with the ratio of Majorana neutrino's virtuality $(\sqrt{p_N^2})$ over its mass. For virtualities larger than $m_N$, there is an enhancement of the $\Delta L = 0$ transition probability over the $\vert \Delta L\vert = 2$ mode, i.e., $\Delta_\mathcal{M}^{\rm Large~Width}>0$. We trace this to the $L$-conserving $(\ell_1^+  N \ell_2^-)$ lepton current, which as shown in Eq.~\eqref{eq:helicityAmp_lnc_reduced}, selects for the LH helicity of $N$ and is thus proportional to its momentum $(\not\!\! p_N)$. In the $L$-violating case, the $(\ell_1^+  N \ell_2^+)$ current selects for the RH helicity of $N$ and thus is proportional to its mass $(m_N \mathbb{I}_4)$. For virtualities smaller than $m_N$, the relative helicity enhancement/suppression is inverted with $\Delta_\mathcal{M}^{\rm Large~Width}<0$.

In an altogether different limit, it may be that $N$ is too heavy to ever be resonantly produced in $W$ boson decays. In this case, one enters the decoupling limit~\cite{Appelquist:1974tg} and the pole structure of $N$'s propagator behaves as
  \begin{align}
  \mathcal{D}(\tilde{p}_k)  &= \cfrac{i}{(\tilde{p}_k^2 - m_N^2) + i(\Gamma_N m_N)}
  \\
  &= \cfrac{-i}{m_N^2}\left[ 1 + \mathcal{O}\left(\frac{\tilde{p}_k^2}{m_N^2}\right)  + i \mathcal{O}\left(\frac{\Gamma_N}{m_N}\right)  \right].
  \label{eq:prop_decouple}
  \end{align}
  Importantly, the propagator's leading contribution is the same for the $\Delta L = 0$ ME as well as both diagrams in the $\vert \Delta L\vert = 2$  ME. Thus, any difference between the two transition rates is ultimately due to helicity inversion.
  
  After propagating this expansion, the squared MEs for the $L$-conserving and $L$-violating $W$ boson decays are:
  \begin{align}
\sum \vert \mathcal{M}_\lnc^W \vert^2 & \sim 
 \cfrac{M_W^2 p_N^2}{m_N^4}
\left[ 1 + \mathcal{O}\left(\frac{p_N^2}{m_N^2}, ~ \frac{\Gamma_N}{m_N}\right) \right], 
\\
\sum \vert \mathcal{M}_\lnv^W \vert^2 &\sim 
  \cfrac{M_W^2}{m_N^2}
\left[ 1 + \mathcal{O}\left(\frac{p_N^2}{m_N^2},  ~ \frac{\Gamma_N}{m_N}\right) \right]. 
\end{align}
Likewise, their difference $(\Delta_\mathcal{M}^{\rm Large~Mass})$ is given by
\begin{align}
&\Delta_\mathcal{M}^{\rm Large~Mass}  \equiv \sum \vert \mathcal{M}_\lnc^W \vert^2 - \sum \vert \mathcal{M}_\lnv^W \vert^2 \\
&\sim 
\cfrac{M_W^2}{m_N^2} 
 \left( \cfrac{p_N^2}{m_N^2} - 1 \right) \times
\left[ 1 + \mathcal{O}\left(\frac{p_N^2}{m_N^2}, ~ \frac{\Gamma_N}{m_N}\right) \right].
 \label{eq:asymm_offshell_largeMass}
\end{align}

Immediately, we see that the $L$-conserving case exhibits a quartic dependence on the Majorana neutrino's mass, whereas the $L$-violating case has only a quadratic dependence.  This reveals  that in the decoupling limit the transition rate for the $\Delta L = 0$ process vanishes faster than the $\vert \Delta L\vert = 2$ transition rate. In the language of effective field theories, this is the manifestation of $L$-conserving operators at dimension eight decoupling more quickly than $L$-violating operators at dimension seven.

As in the large-width scenario, the difference between the two squared MEs stems from the respective preservation and inversion of helicity in the  $(\ell_1^+  N \ell_2^-)$ and $(\ell_1^+  N \ell_2^+)$ lepton currents. More specifically, the $m_N$ factor that is collected in the $\vert \Delta L\vert = 2$ case partially compensates the mass suppression in Eq.~\eqref{eq:prop_decouple}, and reduces the dimension of $L$-violating operators. This is unlike the large-width scenario, where the virtuality of $N$ can exceed its mass and leads to an enhancement of the $L$-conserving transition. The virtuality of $N$ in the decoupling limit is always smaller than its mass and therefore leads to a suppression of the $L$-conserving transition.

\subsection{$2\to4$ scattering with $\Delta L = 0$ and $\vert\Delta  L\vert  = 2$}\label{sec:helicity_scatt}

To extrapolate our  findings, i.e., the existence of helicity inversion but the absence of helicity suppression in $L$-violating decays of $W$ bosons involving (nearly) on-shell Majorana $N$, to other processes, it is helpful to stress that the above arguments are kinematical in nature. They rely on Lorentz invariance, spin correlation, and expansions around leading regions of phase space. They do not rely on strong interference, flavor symmetries, or mixing suppression that one often encounters \cite{Mohapatra:1986aw,Mohapatra:1986bd,Bernabeu:1987gr,Pilaftsis:1991ug,Kersten:2007vk,Antusch:2015mia,Moffat:2017feq}.
As such, the results are process-dependent and are sensitive to whether Eq.~\eqref{eq:massNeighborhood}, or a similar relation, is satisfied.

With this in mind, one direction where it is  possible to extrapolate the above phenomenon is to $2\to n$ scattering processes.  In particular, there is the $L$-conserving, 
\begin{align}
u_L(p_u)\overline{d}_R(p_d) &\to  W^+_{\lambda_W}(p_W) \to \ell^+_{1R}(p_1) N_{\lambda_N}(p_N)   \nonumber\\
 & \to \ell^+_{1R}(p_1) \ell^-_{L2}(p_2) c_L(p_c) \overline{s}_R(p_s), \label{eq:helac_scatt_proc_lnc}
 \end{align}
 as shown  in Fig.~\ref{fig:diagram_HeavyN_DYX_llff}(a),
 and the  $L$-violating,
 \begin{align}
u_L(p_u)\overline{d}_R(p_d) &\to W^+_{\lambda_W}(p_W) \to   \ell^+_{1R}(p_1) N_{\lambda_N}(p_N) \nonumber\\
&   \to \ell^+_{1R}(p_1) \ell^+_{R2}(p_2) \overline{c}_R(p_c) s_L(p_s), \label{eq:helac_scatt_proc_lnv}
\end{align}
as shown  in Fig.~\ref{fig:diagram_HeavyN_DYX_llff}(b). The novelty of these channels follows from the limitations in the $W$ decay  case study. The first limitation relates to the idealization of working with  an unpolarized, on-shell $W$ boson. This is an object that is never really actualized in nature. By virtue of the $W$'s chiral couplings, real $W$s are produced with some degree of polarization~\cite{Ellis:1991qj,Bern:2011ie,Stirling:2012zt}. Likewise, a degree of off-shell virtuality is nearly always present and such contributions are not guaranteed to be negligible if $m_N \sim M_W$~\cite{Berdine:2007uv,Kauer:2007zc,Kauer:2007nt,Uhlemann:2008pm}.

To check the impact of these matters on the existence of inversion and suppression in Eqs.~\eqref{eq:helac_scatt_proc_lnc}-\eqref{eq:helac_scatt_proc_lnv}, we again construct the associated MEs. These can be built from the MEs in Eqs.~\eqref{eq:helicityAmp_lnc_full} and \eqref{eq:helicityAmp_lnv_full} for a $W$ decay to an approximately on-shell $N$ by working in the hard scattering frame with the following momentum assignments
\begin{align}
p_u =\frac{Q}{2}(1,0,0,1), &\quad p_d = \frac{Q}{2}(1,0,0,-1), \\
p_W = p_1+p_2, &\quad Q^2 = p_W^2 = (p_1+p_2)^2.
\end{align}
After substituting the $W$ polarization vector  for the current and propagator of the sub-process $u_L\overline{d}_R\to W^{+*}$,
\begin{equation}
\varepsilon_\mu(p_W) \to \tilde{J}^\sigma(p_u,p_d) \Delta_{\sigma\mu}(p_W = p_u+p_d),
\label{eq:meSubstitution}
\end{equation}
where the propagator $\Delta$ is the same as in Eq.~\eqref{eq:helicityAmp_WProp}
and the $(u_L\overline{d}_R)$ current $ \tilde{J}$ is given by
\begin{equation}
 \tilde{J}^\sigma(p_u,p_d) = 
 \frac{-i g_W \delta^{AB}}{\sqrt{2}} \left[\overline{v}^B_R(p_d)\gamma^\sigma P_L u^A_L(p_u)\right],
\end{equation}
one obtains the following MEs for the $L$-conserving $(\mathcal{M}_\lnc^{(2\to4)})$ and $L$-violating $(\mathcal{M}_\lnv^{(2\to4)})$ scattering processes:
\begin{align}
-i\mathcal{M}_\lnc^{(2\to4)} =& \tilde{J}^\sigma(p_u,p_d) \Delta_{\sigma\mu}(p_u+p_d)
\nonumber\\
& \times ~T_\lnc^{\nu\mu}(p_1,p_2,p_N) \nonumber\\
& \times  \Delta_{\nu \rho}(p_c+p_s) ~J^\rho(p_c,p_s),
 \label{sec:helicity_me_scatt_lnc}
\end{align}
\begin{align}
-i\mathcal{M}_\lnv^{(2\to4)}  = & \tilde{J}^\sigma(p_u,p_d) \Delta_{\sigma\mu}(p_u+p_d)
\nonumber\\
& \times ~T_\lnv^{\nu\mu}(p_1,p_2,p_1+p_c+p_s) \nonumber\\
& \times  \Delta_{\nu \rho}(p_c+p_s) ~J^\rho(p_s,p_c) \nonumber\\
 + & (p_1\leftrightarrow p_2).
 \label{sec:helicity_me_scatt_lnv}
\end{align}

To extract the scaling behavior of these two MEs, we exploit the fact that the $W$'s longitudinal polarization $(\lambda_W=0)$, which generates a different mass-energy power counting than   transverse polarizations $(\lambda_W=\pm1)$, does not couple to massless fermions. It does not contribute to the  $u\overline{d} \to W^* \to N \ell$ sub-process, regardless of external polarizations.  Using this  and after explicit evaluation of the helicity spinor algebra, we obtain for both cases,
\begin{align}
\tilde{J}^\sigma(p_u,p_d) &\Delta_{\sigma\mu}(p_u+p_d) = \nonumber\\
& (-i)^2  \delta^{AB}  \frac{g_W}{\sqrt{2}} 
 \cfrac{\left[\overline{v}^B_R(p_d)\gamma_\mu P_L u^A_L(p_u)\right]}{(Q^2 - M_W^2 + i\Gamma_W M_W)}
 \\
 = &
 (-i)^2  \delta^{AB}  \frac{g_W}{\sqrt{2}} 
 \cfrac{Q\left(0,1,-i,0\right)}{(Q^2 - M_W^2 + i\Gamma_W M_W)}
  \\
 \sim &
\cfrac{Q}{\Gamma_W M_W}\left[ 1 - \mathcal{O}\left(\frac{(Q^2 - M_W^2)}{\Gamma_W M_W}\right) \right].
\end{align}
For clarity, we  expanded the $W$'s propagator in the final line around its on-shell limit, i.e., $(Q^2 - M_W^2)\ll \Gamma_W M_W$.

It is evident that the substitution in Eq.~\eqref{eq:meSubstitution} does not introduce additional parity inversion, say via coupling to longitudinal modes, nor any new dependence on $m_N$. As a consequence, the scaling behavior of the  $(\ell_1^+  N \ell_2^\mp)$ lepton currents and propagators in the scattering process are the same as in the decay process, up to substitutions of the total c.m. energy:  $M_W \to Q$. Consistently, this means external momenta scale as $E_{external} \sim Q$.

Propagating these modifications, one finds that in the double on-shell limit,  the leading contributions to the squared MEs for the $2\to4$ processes  scale as
  \begin{align}
\sum & \vert \mathcal{M}_\lnc^{(2\to4)} \vert^2  \sim 
 \cfrac{Q^4 p_N^2}{(\Gamma_W M_W)^2(\Gamma_N m_N)^2}
 \\
 & \quad \times \left[ 1 - \mathcal{O}\left(\frac{(Q^2 - M_W^2)}{\Gamma_W M_W}\right) - \mathcal{O}\left(\frac{\delta p_N^4}{(\Gamma_N m_N)^2}\right) \right],
 \nonumber\\
 \sum & \vert \mathcal{M}_\lnv^{(2\to4)} \vert^2 \sim 
  \cfrac{Q^4}{(\Gamma_W M_W)^2\Gamma_N^2}
  \\
  & \quad \times \left[ 1 - \mathcal{O}\left(\frac{(Q^2 - M_W^2)}{\Gamma_W M_W}\right) - \mathcal{O}\left(\frac{\delta p_N^4}{(\Gamma_N m_N)^2}\right) \right].
  \nonumber
  \end{align}
  As in the $1\to4$ decays, we find that the helicity inversion in $2\to4$ scattering does not manifest as helicity suppression in the on-shell limit for $N$ when $\Gamma_N\ll m_N$. In fact, we find again that the squared ME for the $L$-conserving and $L$-violating processes are the same, up to the heavy neutrino's off-shellness. Thus, one  obtains equal cross sections in the absence of phase space cuts. This lack of helicity suppression/enhancement for off-shell gauge mediators is consistent with past studies on related phenomena~\cite{Han:2012vk,Ruiz:2017nip,Nemevsek:2018bbt}.  When $N$ is dominantly off-shell, the same enhancements/suppressions described in Sec.~\ref{sec:helicity_offshell} appear, up to appropriate $M_W \to Q$ substitutions.
  
Following analogous arguments, we anticipate that these findings hold also for cascade decay processes, such as  top quark decays to Majorana neutrinos, $t \to b \ell_1 N \to   b \ell_1 \ell_2 f_1 \overline{f_2}$~\cite{BarShalom:2006bv,Si:2008jd,Liu:2019qfa}. In such situations, contributions from the $W$'s longitudinal polarization may introduce additional dependencies on $(Q/M_W)\sim (m_t/M_W)$ but otherwise not alter the tensor structure of the $(\ell_1 N \ell_2)$ lepton  currents. As such, the scaling of squared momenta for $N$ will remain the same in its near on-shell limit.
    
As a brief remark, we note that at next-to-leading order in EW it may be that differences in the $L$-conserving and $L$-violating processes generate asymmetric transition rates. Likewise, while virtual $\mathcal{O}(\alpha_s)$ corrections to the processes in Fig.~\ref{fig:diagram_HeavyN_DYX_llff} will not impact the polarization of the intermediate $N$~\cite{Ruiz:2015zca}, the expectation for real $\mathcal{O}(\alpha_s)$ emissions is less clear. In principle, these effects are coupling-suppressed but such considerations are left for the future.

\subsection{Other Scenarios  with $\vert\Delta L\vert = 2$}\label{sec:helicity_other}

A second direction where one can apply the above findings is to other new physics scenarios that feature chiral gauge interactions  and Majorana fermions. While a systematic survey is beyond the present scope of this work, two concrete examples are: warped extra dimension with RH Majorana neutrinos $(\nu_R)$~\cite{ArkaniHamed:1998vp,Grossman:1999ra,Csaki:2008qq} and the Left-Right Symmetric model (LRSM) \cite{Pati:1974yy,Mohapatra:1974hk,Mohapatra:1974gc,Senjanovic:1975rk,Mohapatra:1979ia,Mohapatra:1980yp}.

The first is characterized by Kaluza-Klein (KK) excitations of SM particles as well as of $\nu_R$. This includes, for example, $W'_{KK}$ gauge bosons, which have the same chiral interaction structure and gauge quantum numbers as the SM $W$ boson. After mass-diagonalization, the resulting Lagrangian that governs interactions between the mass eigenstates $N_{KK}$, $W^{'\pm}_{KK}$, and $\ell^\pm_{KK}$ is essentially the same as Eq.~\eqref{eq:nuLag}, up to an overall rescaling of couplings. 

Phenomenologically speaking, this allows processes  like those shown in Fig.~\ref{fig:diagram_HeavyN_DYX_llff} but with internal particles substituted with their KK excitations. Corresponding MEs and squared MEs are therefore the same as those constructed in Secs.~\ref{sec:helicity_lnc}-\ref{sec:helicity_scatt}, up to substitutions of mass and coupling constants, implying the presence of helicity inversion.  So long as external particles are massless and the near on-shell condition of Eq.~\eqref{eq:massNeighborhood} is satisfied, one should consistently find an absence of helicity suppression, modulo off-shell virtuality and finite width effects.

In  the second case, the LRSM is characterized by embedding the SM's
 $\mathcal{G}_{SM} = SU(3)_c \otimes SU(2)_L \otimes U(1)_Y$ gauge symmetry into the larger symmetry group, 
 $ \mathcal{G}_{LRSM} = SU(3)_c \otimes SU(2)_L \otimes SU(2)_R \otimes U(1)_{B-L}\otimes \mathcal{P}$.
In  this model, all of the SM's RH  chiral fields and $\nu_R$ are charged under the $SU(2)_R$ gauge group, just as their LH counter parts are charged under  $SU(2)_L$. The $U(1)$ conservation of baryon-minus-lepton  numbers $(B-L)$ ensures that the theory is anomaly free and the generalized discrete parity $\mathcal{P}$ ensures that the LH and RH gauge interactions are identical before spontaneous symmetry breaking.

After LR and EW symmetry breaking, one finds RH gauge bosons $W_R$ that couple to heavy Majorana neutrinos $N$ and charged leptons $\ell$ through RH chiral currents, in analogy to the Lagrangian in Eq.~\eqref{eq:nuLag}.
This leads  to the spectacular $L$-violating scattering process~\cite{Keung:1983uu}
 \begin{align}
u_R\overline{d}_L ~\to~ W^+_R  ~\to~  \ell^+_{1L} N   
&\to~ \ell^+_{1L} \ell^+_{L2} W_R^{-*} \nonumber\\
&\to~ \ell^+_{1L} \ell^+_{L2} \overline{c}_L s_R.
\end{align}
This associated diagram is the same as Fig.~\ref{fig:diagram_HeavyN_DYX_llff_LNV}  but with substituting the SM gauge boson $W$ for LRSM gauge boson $W_R$. Explicit computation reveals a ME that is identical to the $L$-violating ME of Eq.~\ref{sec:helicity_me_scatt_lnv}, up to substitutions of masses and couplings as well as the exchange of $P_L$ chiral projection operators for the operator $P_R$.

Importantly, the consistent application of the Feynman rules of Ref.~\cite{Denner:1992vza,Denner:1992me} requires the vertex modifications
\begin{align}
\gamma^\nu P_R &\to
(-1) \gamma^\nu P_L
\\
\overline{v}_L(p_2) &\to \overline{u}_L(p_2). 
\label{eq:feynmanrule_03}
\end{align}
This leads to an explicit helicity inversion at the ME level as in  the Phenomenological Type I Seesaw. Assuming that the near on-shell condition for intermediate resonances is  satisfied, one again finds that the explicit dependence on $m_N^2$ cancels. Again, this leads to an absence of helicity suppression, up to the now-usual caveats.

For completeness, one could also consider the mixed $W_R-W_L$ scattering process given by~\cite{Han:2012vk}
 \begin{align}
u_R\overline{d}_L ~\to~ W^+_R  ~\to~  \ell^+_{1L} N   
&\to~ \ell^+_{1L} \ell^+_{R2} W_L^{-*} \nonumber\\
&\to~ \ell^+_{1L} \ell^+_{R2} \overline{c}_R s_L.
\end{align}
In this case, one finds a \textit{second} helicity inversion due to inverting  the chiral coupling associated with the second charged current. This implies that the roles are now reversed: the $L$-violating process exhibits a net helicity conservation while the $L$-conserving process exhibits a net helicity inversion. Explicit calculation~\cite{Han:2012vk} again shows a lack of helicity suppression in the near on-shell limit.

\section{Numerical impact  of helicity inversion in $\vert \Delta L\vert=2$ LHC processes}\label{sec:results}

In light of the previous section, the question is not  whether there is helicity inversion in $\vert \Delta L \vert = 2$ amplitudes mediated by Majorana neutrinos in the Phenomenological Type I Seesaw. It exists and follows from a parity inversion in EW interactions. The pertinent issue is whether contributions from off-shell virtualities, which can give rise to helicity-suppressing behavior, is numerically relevant for searches for Majorana $N$ at the LHC.

To investigate this, we consider two complementary measures of helicity suppression. The first, presented in section~\ref{sec:results_width}, is based the potential asymmetry that could develop in $L$-conserving and $L$-violating decays of the SM $W$ boson. The second, presented in section~\ref{sec:results_xsec}, is the analogous asymmetry that can appear in hadronic $2\to4$ cross sections. For both cases we inherently work in a limit where resonant production of $N$ dominates. Therefore in section~\ref{sec:results_offshell} we investigate the possible importance of off-shell contributions. Before presenting our numerical results, we comment in Sec.~\ref{sec:results_spin} on the preservation of spin-correlation in our computations and then validate the presence of strong helicity inversion in Sec.~\ref{sec:results_ratio}.

\subsection{Numerical preservation of spin-correlation}\label{sec:results_spin}

To undertake our numerical computations we  exploit the massive spinor helicity formalism of Refs.~\cite{Hagiwara:1985yu,Hagiwara:1986vm,Hagiwara:1988pp,Murayama:1992gi}  as implemented in the \texttt{ALOHA} package~\cite{deAquino:2011ub,Alwall:2014hca}, in the \texttt{HELAS}  basis~\cite{Hagiwara:1985yu}. (For precise details of the computational setup, see section~\ref{sec:setup}.) We do so in order to evaluate MEs exactly but at the cost of analytical expressions.

We forego analytical expressions due to the fact that we are dealing with multi-scale, $1\to4$ and $2\to4$ processes. The squared MEs for these processes must be amended with  kinematic factors and integrated over phase space to derive total decay widths $(\Gamma)$ and cross sections $(\sigma)$, i.e., the  quantities considered here. In the absence of strong assumptions like the narrow width approximation (NWA), phase space integration usually leaves intractable  algebraic expressions for such processes.  However, we avoid employing the NWA since its rigorous justification  for  EW-scale Majorana  neutrinos  is not well-established in the literature.  On the contrary,  studies into the validity of the NWA itself list criteria  that may not be satisfied here~\cite{Berdine:2007uv,Kauer:2007zc,Kauer:2007nt,Uhlemann:2008pm,Artoisenet:2012st}, and even show~\cite{Berdine:2007uv} a sizable impact on the spin-correlation propagated by Majorana fermions. While important, such  considerations are  outside our scope and deferred to later  work.

\begin{table}[t!]
\begin{center}
\resizebox{.65\columnwidth}{!}{
\begin{tabular}{rcc||c}
\hline\hline
\multicolumn{4}{c}{$W^+(\lambda_W) ~\to~ e^+(\lambda_{e}) ~  N(\lambda_N)   $} \\
\hline
$\lambda_W$	& $\lambda_e$	&	$\lambda_N$		& 	$-i\mathcal{M}(\lambda_W, \lambda_e, \lambda_N)  ~/~  \left(\frac{-i g_W}{\sqrt{2}}V_{e N}\right) $ 				
\\ 
\hline
$+1$	& $R$	& $R$	& $\frac{1}{\sqrt{2}}m_N\sqrt{1-r_N} \sin\theta_e $					\\ \hline
$0$	& $R$	& $R$	& $-m_N\sqrt{1-r_N} \cos\theta_e e^{-i\phi_e}$					\\ \hline
$-1$	& $R$	& $R$	& $-\frac{1}{\sqrt{2}}m_N\sqrt{1-r_N} \sin\theta_e e^{-i2\phi_e}$		\\ \hline
$+1$	& $R$	& $L$	& $\frac{1}{\sqrt{2}}M_W\sqrt{1-r_N} (1+\cos\theta_e) e^{i\phi_e}$ 	\\ \hline
$0$	& $R$	& $L$	& $M_W\sqrt{1-r_N} \sin\theta_e$								\\ \hline
$-1$	& $R$	& $L$	& $\frac{1}{\sqrt{2}}M_W\sqrt{1-r_N} (1-\cos\theta_e) e^{-i\phi_e}$	\\ \hline
All	& $L$	& All	& 	$0$	\\ \hline
\hline
\end{tabular}
} 
\caption{Helicity amplitudes for the $W^+(\lambda_W) \to e^+(\lambda_{e})  N(\lambda_N) $ decay, with kinematics defined in Sec.~\ref{sec:results_ratio}.
}
\label{tab:decayME}
\end{center}
\end{table}

\subsection{Numerical validation of helicity inversion} \label{sec:results_ratio}

As a first step to quantifying potential helicity suppression in $\vert\Delta L\vert=2$ transitions, we move to establish that our computational setup captures the helicity inversion in such processes. To demonstrate this and in the notation of Sec.~\ref{sec:helicityTheory} we consider the simpler $1\to2$ decay
\begin{equation}
W^+_{\lambda_W}(p_W) ~\to~ e^+_{\lambda_{e}}(p_e)  ~ N_{\lambda_N}(p_N).
\end{equation}
In the $W$  boson's rest  frame and with the assignments,
\begin{align}
p_e =& E_e (1, \sin\theta_e \cos\phi_e, \sin\theta_e\sin\phi_e, \cos\theta_e), \\
p_W =& M_W(1,0,0,0), \quad E_e = \frac{M_W}{2}(1-r_N),\\
p_N 	=& p_W - p_e, \quad  r_N \equiv \left(\frac{m_N}{M_W}\right)^2,
\end{align}
we evaluate and report the amplitude $\mathcal{M}(\lambda_W, \lambda_e, \lambda_N)$ for each helicity permutation $(\lambda_W,\lambda_e,\lambda_N)$ in Tab.~\ref{tab:decayME}.

Several notable features can be identified in the MEs of Tab.~\ref{tab:decayME}. First is that all amplitudes for $e^+(\lambda_e=L)$ are zero, which  is consistent with $W$ bosons only coupling to massless LH particles (RH antiparticles).  Second  is that amplitudes for $\lambda_W=\pm1$ and $\lambda_N=L$ feature the characteristic $(1\pm\cos\theta)$ behavior  associated with vector currents. Third, and  most  relevant, is that amplitudes for $\lambda_N = R$ scale with the  mass of $N$, i.e., $-i\mathcal{M}\sim m_N$, whereas amplitudes for $\lambda_N = L$ scale with the energy of $N$, i.e., $ -i\mathcal{M}\sim E_N \sim M_W$, as one would expect for helicity inversion  of massive decay products.

\begin{table}[t]
\begin{center}
\resizebox{.7\columnwidth}{!}{
\begin{tabular}{rcc||c}
\hline\hline
\multicolumn{4}{c}{$W^+(\lambda_W) ~\to~ e^+(\lambda_{e})  N(\lambda_N)   $} \\
\hline
$\lambda_W$	& $\lambda_e$	&	$\lambda_N$		& 	$\Gamma(\lambda_W, \lambda_e, \lambda_N)   $ 				
\\ 
\hline
$+1$	& $R$	& $R$	& $\frac{g_W^2  }{96\pi} \vert V_{e N}\vert^2  m_N \left(\frac{m_N}{M_W}\right)   (1-r_N)^2  $	\\ \hline
$0$	& $R$	& $R$	& $=\Gamma(+, R, R)$					\\ \hline
$-1$	& $R$	& $R$	& $=\Gamma(+, R, R)$		\\ \hline
$+1$	& $R$	& $L$	& $\frac{g_W^2  }{48\pi} \vert V_{e N}\vert^2  M_W   (1-r_N)^2 $ 	\\ \hline
$0$	& $R$	& $L$	& $=\Gamma(+, R, L)$								\\ \hline
$-1$	& $R$	& $L$	& $=\Gamma(+, R, L)$	\\ \hline
All	& $L$	& All	& 	$0$	\\ \hline
\hline
\end{tabular}
} 
\caption{
Same as Tab.~\ref{tab:decayME}  but for the $W^+(\lambda_W) ~\to~ e^+(\lambda_{e})  N(\lambda_N) $ partial width.
Note: the spin-averaging factor of $\mathcal{S}_W =3$ is not included  in $\Gamma$.
}
\label{tab:decayGam}
\end{center}
\end{table}

Using the definition of the partial decay width for the unpolarized  particle $\mathcal{B}$ with mass $m_{\mathcal{B}}$ into final-state $f$,
\begin{align}
\Gamma(\mathcal{B} \to f_{\lambda_f})  &=  \frac{1}{2m_\mathcal{B} \mathcal{S}_\mathcal{B} N_c^{\mathcal{B}}}  \nonumber\\
 \times& \int d PS_f \sum_{\rm dof} ~ 
\Big\vert\mathcal{M}(\mathcal{B}_{\lambda_{\mathcal{B}}} \to f_{\lambda_f})\Big\vert^2,
\label{eq:widthDef}
\end{align}
we report in Tab. \ref{tab:decayGam} the partial width  $\Gamma(\lambda_W, \lambda_e, \lambda_N)$ for each permutation of helicities $(\lambda_W,\lambda_e,\lambda_N)$.
We  note that, for consistency, the spin-averaging factor of $\mathcal{S}_W=3$ is not included  in $\Gamma(\lambda_W, \lambda_e, \lambda_N)$. This implies  that the canonical spin-averaged total is related by
\begin{equation}
\Gamma\left(W^+\to e^+ N\right) = 
\frac{1}{\mathcal{S}_W }\sum_{{\lambda_k}}\Gamma\left(W^+_{\lambda_W}\to e^+_{\lambda_e} N_{\lambda_N}\right)
\end{equation}
Likewise, the partial and total widths of $W$ are related to its branching rate (BR) by the usual definition
\begin{equation}
{\rm BR}(W\to f) \equiv \cfrac{\Gamma(W\to f)}{\Gamma_W} = \cfrac{\Gamma(W\to f)}{\sum_X \Gamma(W\to X)}.
\end{equation}

In comparison to the MEs, we observe in the partial widths listed in Tab.~\ref{tab:decayGam} that several kinematic features are washed out after phase space integration. In particular, the characteristic $(1\pm\cos\theta)$ behavior  and sensitivity to the azimuthal angle $\phi_e$ are no longer manifest. What remains, however, is the relative dependence on the heavy neutrino's mass.  For the $\lambda_N=L$ cases, we see that the ME's linear power dependence on $M_W$ remains linear in the partial widths. The quadratic power one obtains at the squared ME level is canceled by the explicit mass factor in the definition of $\Gamma$ in Eq.~\eqref{eq:widthDef}. For the $\lambda_N=R$ cases, the linear power dependence on $m_N$ at the ME level grows at the squared ME level, and leads the partial widths to scale as $\Gamma \sim m_N (m_N/M_W)$. Interestingly, this shows that in the fixed $m_N$ but large $M_W$ limit, the MEs for $\lambda_N=R$ marginally grow and converge, whereas the partial widths vanish.
 This behavior is consistent with expectations from the Confusion Theorem.

In taking the ratio of the $W^+\to e^+ N_{\lambda_N}$ branching rates, we can extract the helicity suppression of $\lambda_N=R$ helicity states at small $(m_N/M_W)^2$, and verify the modeling in our setup. Analytically the ratio  is given by
\begin{align}
\mathcal{R}&\equiv
\cfrac{{\rm BR}\left(W^+\to e^+_R N_R\right)}{{\rm BR}\left(W^+\to e^+_R N_L\right)} 
= \cfrac{\Gamma\left(W^+\to e^+_R N_R\right)}{\Gamma\left(W^+\to e^+_R N_L\right)} 
\\
& = \cfrac{\frac{1}{\mathcal{S}_W}\sum_{\lambda}\Gamma\left(W^+_{\lambda}\to e^+_R N_R\right)}{\frac{1}{\mathcal{S}_W}\sum_{\lambda}\Gamma\left(W^+_{\lambda}\to e^+_R N_L\right)}  
\\
&=\frac{1}{2}\left(\frac{m_N}{M_W}\right)^2.
\end{align}

\begin{figure}
\includegraphics[width=0.46\textwidth]{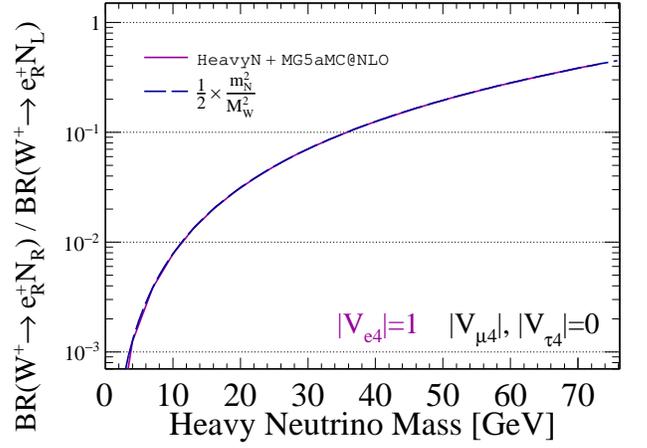}  
\caption{
The $W^+\to e^+_R N_R$ and $W^+\to e^+_R N_L$  branching ratio 
as a function of $m_N$ [GeV], 
 as computed numerically from polarized matrix elements (solid) and analytically (dashed).
}
\label{fig:lnvHelic_SM_WL_BRratio}
\end{figure}

In Fig.~\ref{fig:lnvHelic_SM_WL_BRratio} we plot $\mathcal{R}$ as a function of heavy neutrino mass $m_N$ [GeV] as computed numerically from polarized matrix elements (solid line) and analytically (dashed line). For heavy neutrino masses in the range of $m_N\in[1\GeV, 75\GeV]$ we find that $\mathcal{R}$ spans 3-4 orders of magnitude. Over this entire range we find excellent agreement between our numerical setup and exact analytic expectations. This provides nontrivial checks that (i) helicity inversion for viable values of heavy neutrino masses can be numerically significant, and (ii) our computational setup successfully captures such behavior.

Briefly, we note that we do not consider Majorana neutrinos with masses below $m_N = 1\GeV$. For such states the relevant virtuality scales are comparable to the non-perturbative scale of QCD. Hence, one should treat the decays  of lighter sterile neutrinos, i.e., for $m_{N_k}\lesssim 1-10\GeV$, like decays of $\tau$ leptons and adopt a low-energy, effective field theory, as done for example  in Refs.~\cite{Atre:2009rg,Balantekin:2018ukw,Coloma:2020lgy}. This introduces additional parity  nuances that have been considered elsewhere~\cite{Balantekin:2018ukw}.

\begin{figure*}
\subfigure[]{\includegraphics[width=0.46\textwidth]{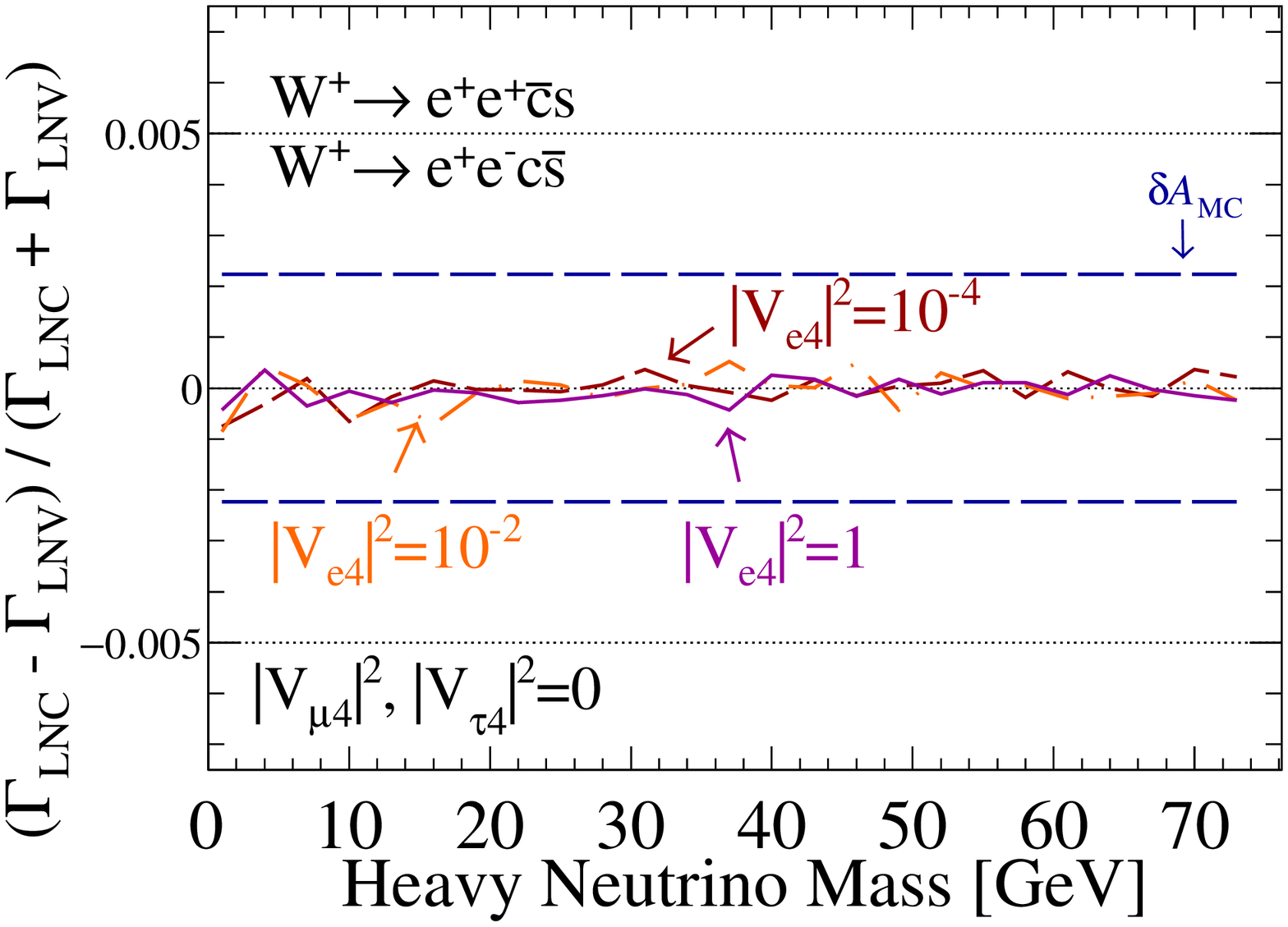}	\label{fig:lnvHelicity_SM_WL_WidthAsymm}}
\subfigure[]{\includegraphics[width=0.46\textwidth]{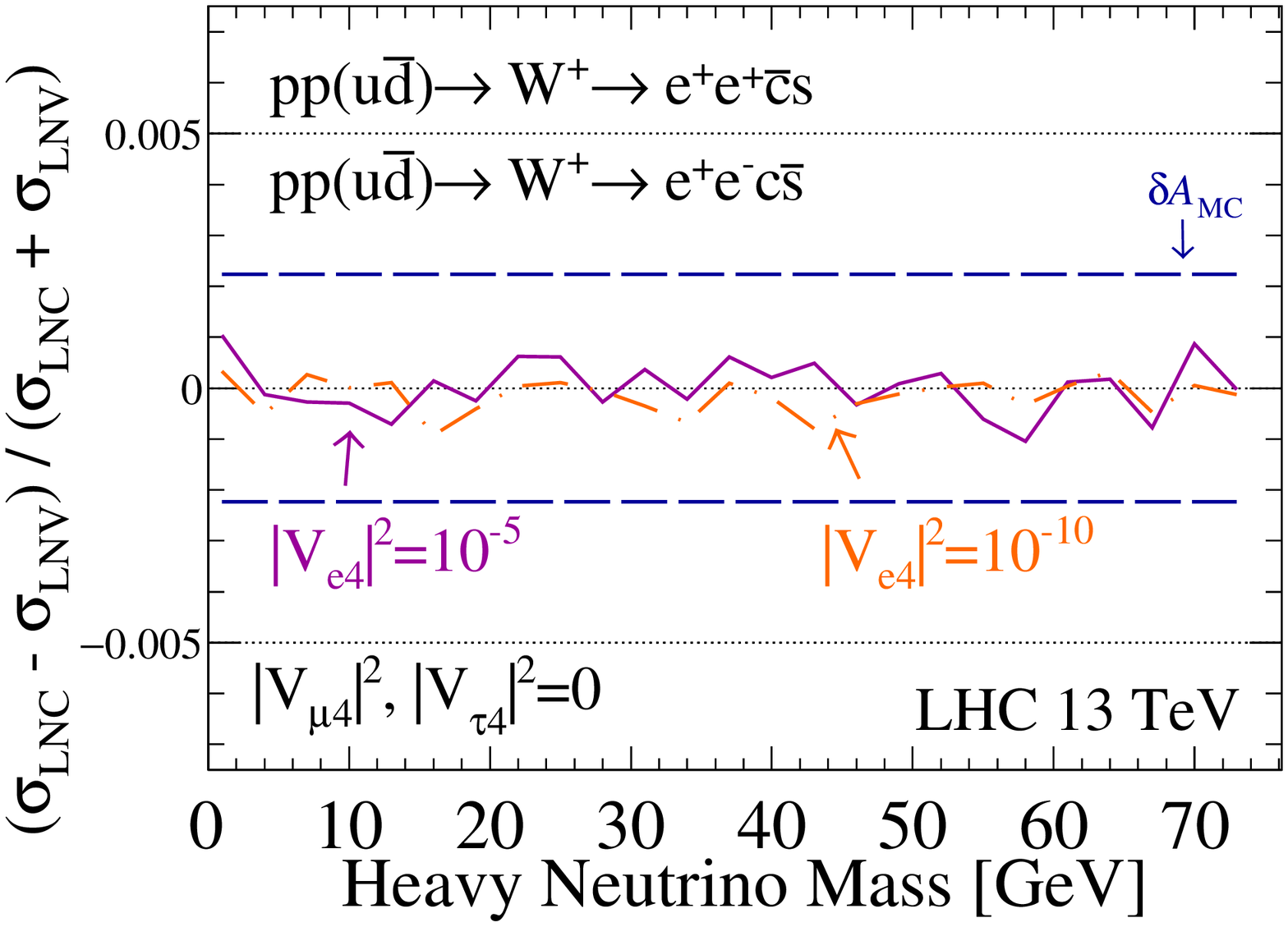}		\label{fig:lnvHelic_SM_WL_XSecAsymm}}
\caption{
(a) Decay rate asymmetry between the $L$-conserving $W^+ \to e^+_1 N  \to e^+_1 e_2^- c \overline{s}$ $(\Gamma_{\rm LNC})$ and $L$-violating $W^+ \to e^+_1 N  \to e^+_1 e_2^+ \overline{c} s$  $(\Gamma_{\rm LNV})$ processes as a function of heavy neutrino mass $m_N$ [GeV] for representative active-sterile neutrino mixing $\vert V_{e N}\vert$.
(b) Same, but for the hadron-level,  cross section asymmetry between the $L$-conserving $pp(u\overline{d}) \to e^+_1 N  \to e^+_1 e_2^- c \overline{s}$ $(\sigma_{\rm LNC})$ and  $L$-violating $pp(u\overline{d}) \to e^+_1 e_2^+ \overline{c} s$ $(\sigma_{\rm LNV})$ processes at $\sqrt{s}=13\TeV$.
Also shown are statistical MC uncertainty bands ($\delta A_{\rm MC}$).
}
\label{fig:lnx_asymmetry}
\end{figure*}

\subsection{Total Width  Asymmetry}\label{sec:results_width}

As our first measure of helicity suppression in LHC observables for processes that are mediated by heavy Majorana neutrinos, we consider respectively the $L$-conserving and $L$-violating,  $1\to4$-body $W$ boson decay processes,
\begin{align}
\Delta L = 0 		~: &~ W^+ \to e^+_1 N  \to e^+_1 e_2^- c \overline{s}, \label{eq:decay_proc_lnc}\\
\vert \Delta L\vert = 2 ~: &~ W^+ \to e^+_1 N  \to e^+_1 e_2^+ \overline{c} s. \label{eq:decay_proc_lnv}
\end{align}
Here we fix final-state flavors  for definiteness. Diagrams with $\gamma^*/Z^*$ exchange are removed in a gauge-invariant manner, resulting in those shown in Fig.~\ref{fig:diagram_HeavyN_DYX_llff}. Interfering diagrams from identical particle exchange are kept.

In Sec.~\ref{sec:helicityTheory}, we argued  that the ME for these processes exhibit different parametric dependencies on $m_N$ due helicity inversion. At the same time we showed in Sec.~\ref{sec:results_ratio} that Lorentz invariance lead to the same parametric dependence in squared MEs, in the on-shell limit for $N$. Differences in decay rates were found to be proportional to the off-shell virtuality of $N$ as well as to its total width. To address the importance of these terms and quantify the existence of any such helicity suppression, we consider the following asymmetry $\mathcal{A}_\Gamma$ in branching rates:
\begin{align}
\mathcal{A}_\Gamma 
& \equiv \cfrac{{\rm BR}(W^+ \to e^+ e^- c\overline{s})-{\rm BR}(W^+ \to e^+ e^+ \overline{c}{s})}{{\rm BR}(W^+ \to e^+ e^- c\overline{s})+{\rm BR}(W^+ \to e^+ e^+ \overline{c}{s})} 
\\
& = \cfrac{\Gamma(W^+ \to e^+ e^- c\overline{s})-\Gamma(W^+ \to e^+ e^+ \overline{c}{s})}{\Gamma(W^+ \to e^+ e^- c\overline{s})+\Gamma(W^+ \to e^+ e^+ \overline{c}{s})}
\\
& \equiv \cfrac{\Gamma_{\rm LNC}-\Gamma_{\rm LNV}}{\Gamma_{\rm LNC}+\Gamma_{\rm LNV}}.
\label{eq:def_asymm_decay}
\end{align}

In Fig.~\ref{fig:lnvHelicity_SM_WL_WidthAsymm} we show the decay rate asymmetry $\mathcal{A}_\Gamma$  between the $L$-conserving and $L$-violating $W^+$ boson decays given in Eqs.~\eqref{eq:decay_proc_lnc}-\eqref{eq:decay_proc_lnv},  as a function of $m_N$ [GeV] for representative active-sterile neutrino mixing $\vert V_{e N}\vert^2=1$ (solid), $10^{-2}$ (dash-dot), and $10^{-4}$ (dash). Also shown is the associated statistical MC uncertainty band $(\delta \mathcal{A}_{\rm MC})$. Based on $N=100k$ events per determination of $\Gamma$ we obtain a statistical MC uncertainty that is nearly uniform and is approximately $\delta\mathcal{A}_{\rm MC} \approx 2.2\times10^{-3}$.

For heavy neutrino masses in the range of $m_N\in[1\GeV, 75\GeV]$ we report asymmetries consistent with $\mathcal{A}_\Gamma = 0$, i.e., no asymmetry and hence no helicity suppression. More precisely, we find nonzero $\mathcal{A}_\Gamma$ that fluctuate above and below zero, reaching  at most $\vert\mathcal{A}_\Gamma\vert \sim \mathcal{O}(1
\times10^{-3})$, and are consistent with random, statistical noise\footnote{While we use the uncertainty estimator $\delta\mathcal{O}/\mathcal{O}=\delta N/N = 1/\sqrt{N}$, for $\mathcal{O}=\Gamma, \sigma$, it is actually an upper limit on the MC uncertainty due to the sampling and reweighting routines in \texttt{mgamc}~\cite{Alwall:2014hca}.}.
We find that the same behavior holds for all representative choices of active-sterile mixing.
This mixing-independent behavior follows from the definition of $\mathcal{A}_\Gamma$ in Eq.~\eqref{eq:def_asymm_decay}, which indicates that normalization factors of $\vert V_{e N}\vert^2$ in $\Gamma_{\rm LNC}$  and $\Gamma_{\rm LNV}$ cancel in the asymmetry measure.  While the absence of a significant asymmetry seems to suggest a total absence of helicity suppression between  $\Delta L = 0$ and $\vert\Delta L\vert = 2$ decays of $W$ bosons, we stress that the above processes are dominated by regions of phase space where a heavy neutrino is on or is nearly on its mass shell.  As discussed in Sec.~\ref{sec:helicity_offshell}, a different behavior is anticipated outside this limit.

\subsection{Total Cross Section  Asymmetry}\label{sec:results_xsec}

As our second measure of helicity suppression in LHC observables, we consider the generalization  of the $W$ boson decay chains in Eqs.~\eqref{eq:decay_proc_lnc}-\eqref{eq:decay_proc_lnv}. In particular, we consider the $2\to4$-body scattering processes,
\begin{align}
\Delta L = 0 		~: &~ u\overline{d} \to  W^+ \to e^+_1 N  \to e^+_1 e_2^- c \overline{s}, \label{eq:scatt_proc_lnc}\\
\vert \Delta L\vert = 2 ~: &~ u\overline{d} \to W^+ \to e^+_1 N  \to e^+_1 e_2^+ \overline{c} s. \label{eq:scatt_proc_lnv}
\end{align}
We again fix external particle flavors for definiteness and to also avoid interference with the $WW$ scattering process. Diagrams involving  $\gamma^*/Z^*$  exchange are  removed in a gauge-invariant manner, while interfering diagrams from identical particle exchange are kept. As discussed in Sec.~\ref{sec:helicity_scatt}, the utility of these processes is that they capture  polarization and virtuality effects present in real LHC collisions but not in the idealized decays of Sec.~\ref{sec:results_width}.

\begin{figure*}
\subfigure[]{\includegraphics[width=0.46\textwidth]{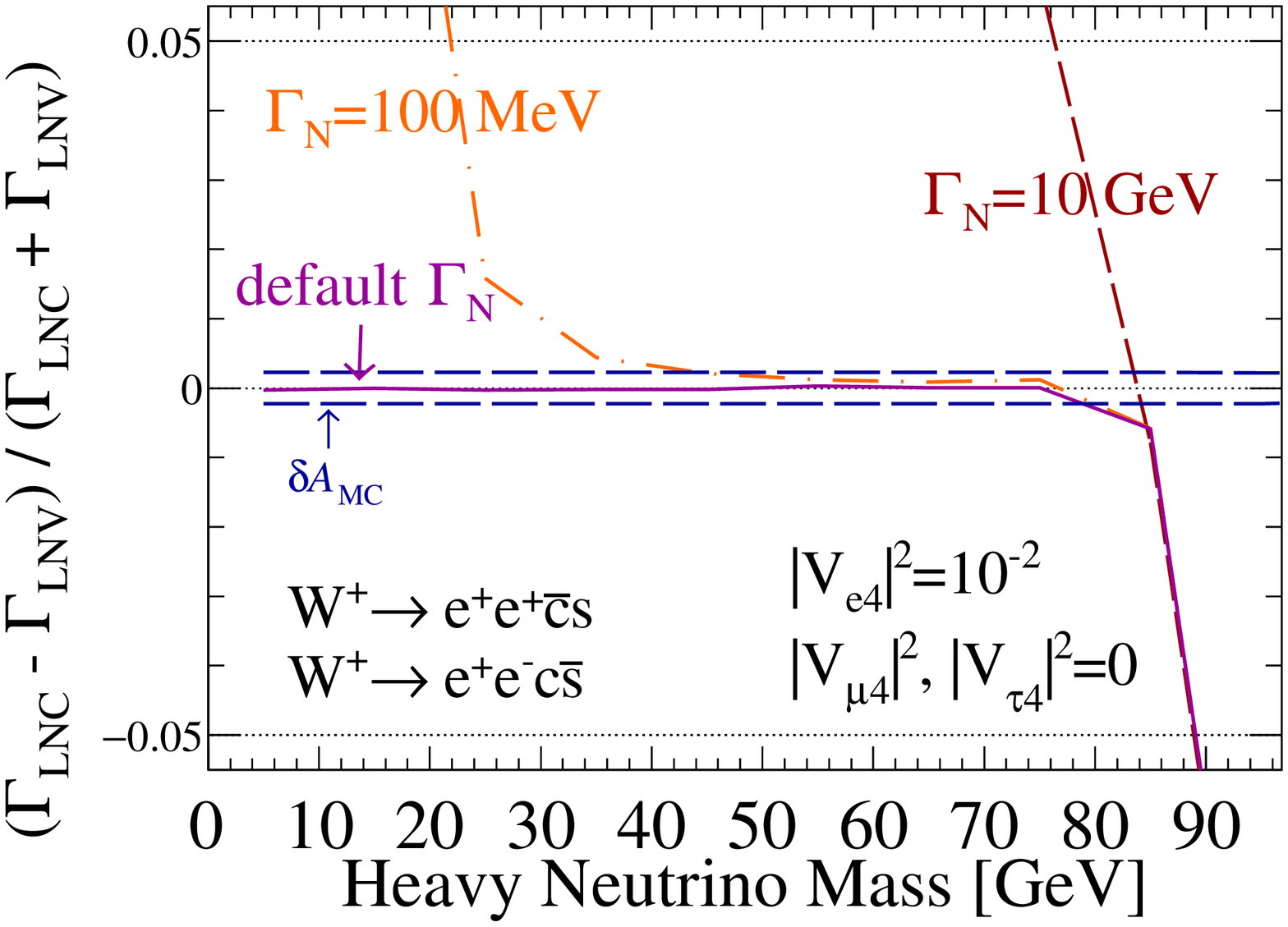}	\label{fig:lnvHelicity_SM_WL_WidthAsymm_LargeWidth_LoMass}}
\subfigure[]{\includegraphics[width=0.46\textwidth]{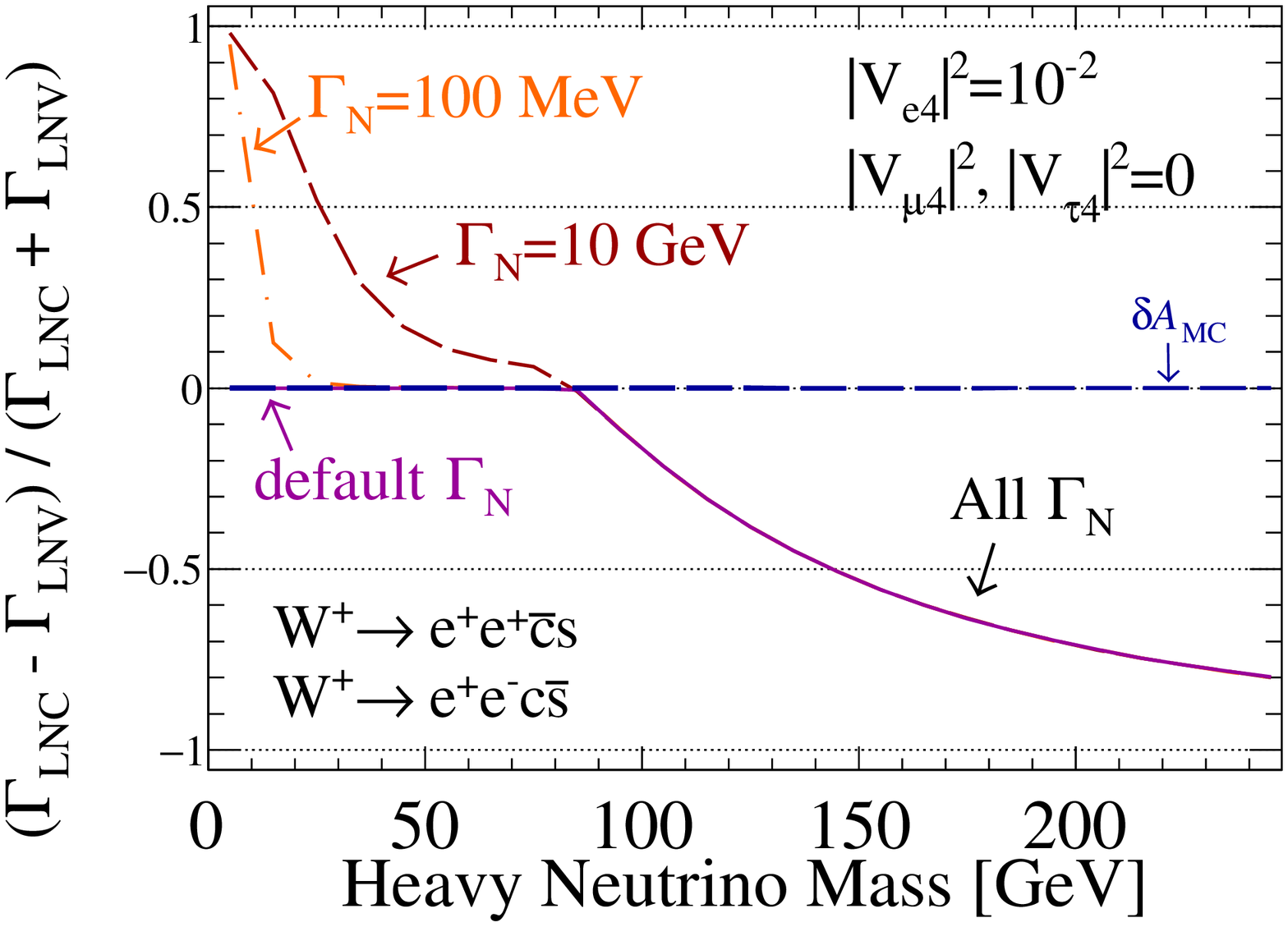}		\label{fig:lnvHelicity_SM_WL_WidthAsymm_LargeWidth_HiMass}}
\\
\subfigure[]{\includegraphics[width=0.46\textwidth]{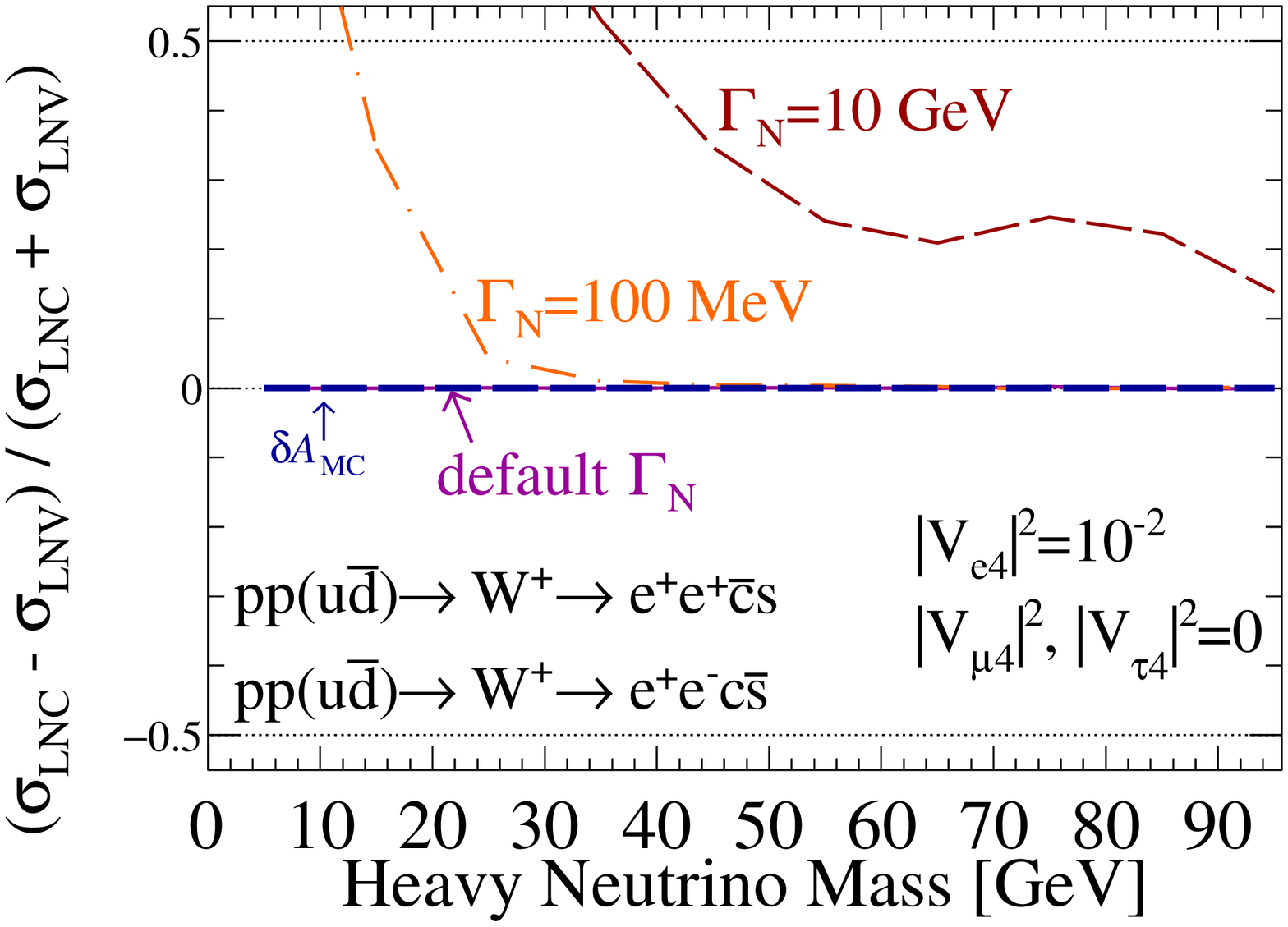}\label{fig:lnvHelicity_SM_WL_XSecAsymm_LargeWidth_LoMass}}
\subfigure[]{\includegraphics[width=0.46\textwidth]{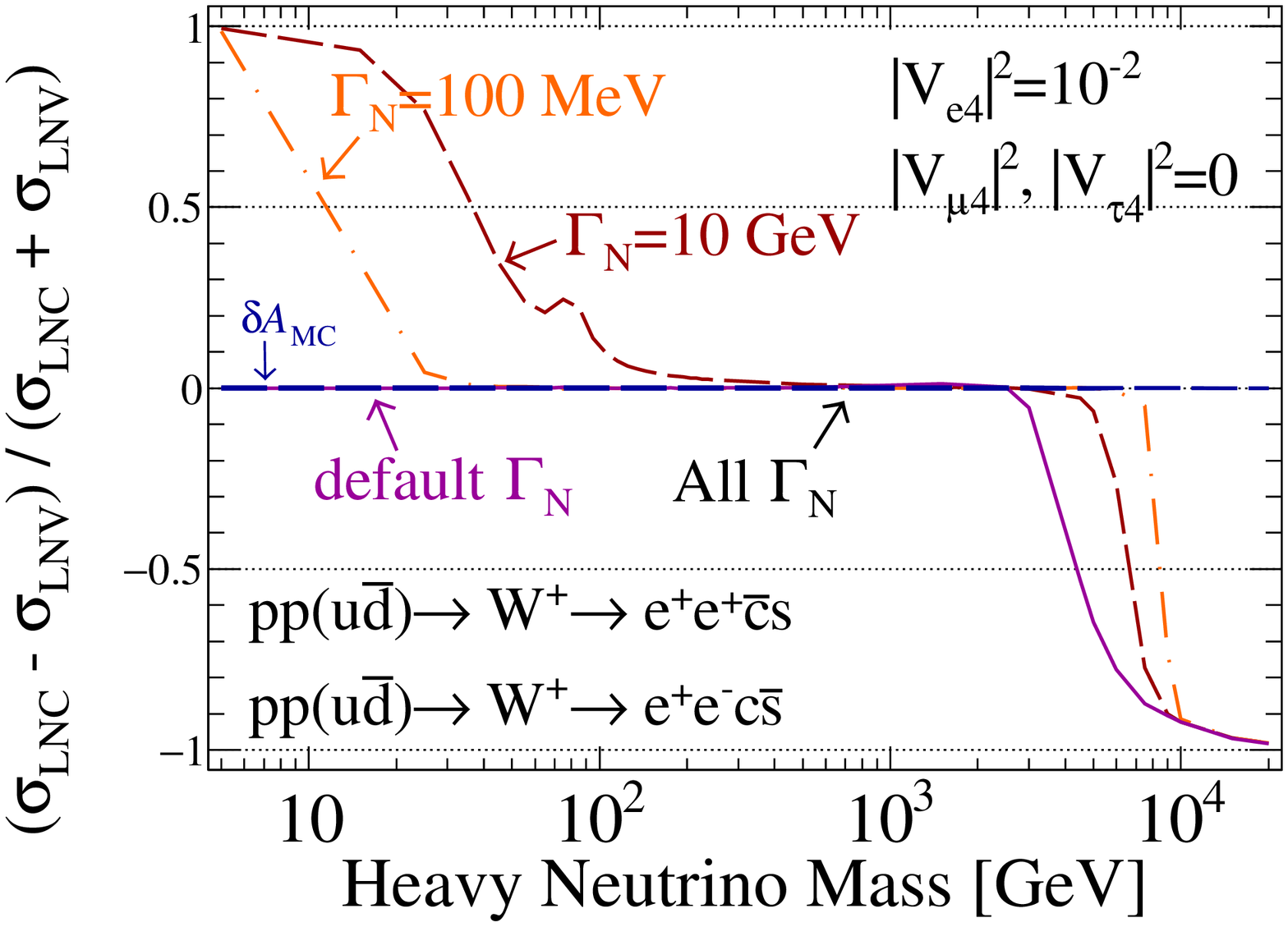}\label{fig:lnvHelicity_SM_WL_XSecAsymm_LargeWidth_HiMass}}
\caption{
(a) Same as Fig.~\ref{fig:lnvHelicity_SM_WL_WidthAsymm}
but for a total width $\Gamma_N$ that varies according to the Lagrangian given in Sec.~\ref{sec:theory_model} (default)
as well as at fixed widths of $\Gamma_N=100\MeV$ and $10\GeV$. For all computations, $\vert V_{e4}\vert^{2}=10^{-2}$ is assumed with $\vert V_{\mu 4}\vert^2, \vert V_{\tau 4}\vert^2 = 0$.
(b) Same as (a) but for an extended mass range.
(c,d) Same as Fig.~\ref{fig:lnvHelic_SM_WL_XSecAsymm} but for the assumptions of (a,b) respectively.
}
\label{fig:lnx_asymmetry_offshell}
\end{figure*}

In analogy to $\mathcal{A}_\Gamma$, we use the scattering processes above to build an asymmetry $\mathcal{A}_\sigma$ that would arise if helicity suppression were to exist. Specifically, we consider 
\begin{align}
\mathcal{A}_\sigma 
& \equiv \cfrac{\sigma(u\overline{d} \to e^+ e^- c\overline{s})-\sigma(u\overline{d} \to e^+ e^+ \overline{c}{s})}{\sigma(u\overline{d} \to e^+ e^- c\overline{s})+\sigma(u\overline{d} \to e^+ e^+ \overline{c}{s})}
\\
&\equiv \cfrac{\sigma_{\rm LNC}-\sigma_{\rm LNV}}{\sigma_{\rm LNC}+\sigma_{\rm LNV}}.
\label{eq:def_asymm_xsec}
\end{align}
Here we abuse slightly the conventional notation for hadronic cross sections $\sigma(pp\to \mathcal{B})$ and write explicitly,
\begin{equation}
\sigma(u\overline{d}  \to \mathcal{B}) = f_{u/p}\otimes f_{\overline{d}/p} \otimes \hat{\sigma}(u\overline{d}   \to \mathcal{B}),
\end{equation}
to denote that we consider only the $u\overline{d}$ partonic contribution to $pp$ scattering, with $f_{i/p}$ representing  the PDF for parton $i$ in hadron $p$, and $\hat{\sigma}$ as the parton-level scattering rate. This is given by the standard expression,
\begin{align}
\hat{\sigma}(ij \to \mathcal{B} )  = &
\frac{1}{2Q^2   ~\mathcal{S}_i \mathcal{S}_j ~N_c^iN_c^j}  \nonumber\\
& \times \int d PS_{\mathcal{B}} \sum_{\rm dof} ~ 
\Big\vert\mathcal{M}(ij\to \mathcal{B})\Big\vert^2.
\end{align}
To avoid potential washout from beam symmetrization, we do not consider the  $\overline{d}u$ partonic channel. 

In Fig.~\ref{fig:lnvHelic_SM_WL_XSecAsymm} we show the cross section asymmetry $\mathcal{A}_\sigma$  between the $L$-conserving and $L$-violating scattering processes in Eqs.~\eqref{eq:scatt_proc_lnc}-\eqref{eq:scatt_proc_lnv},  as a function of $m_N$ [GeV] for representative active-sterile neutrino mixing $\vert V_{e N}\vert^2=10^{-6}$ (solid) and $10^{-10}$ (dash-dot). Also shown is the associated statistical MC uncertainty band $(\delta \mathcal{A}_{\rm MC})$. Remarkably, for heavy neutrino masses in the range of $m_N\in[1\GeV, 75\GeV]$ we report asymmetries that are statistically consistent  with $\mathcal{A}_\sigma = 0$, i.e., no asymmetry and hence. We find that the same behavior holds for both representative choices of active-sterile mixing.

\subsection{Off-shell Effects}\label{sec:results_offshell}

As a final comment, we consider briefly the role of far  off-shell effects in $\Delta L = 0 $ and $\vert \Delta L \vert=2$ decays and cross sections through on-  and off-shell  $W$ bosons. As cautioned above, the $1\to4$ and $2\to4$ transitions that we have so far investigate are dominated by regions of phase space where the intermediate neutrino is (nearly) on-shell. However, as discussed analytically in Sec.~\ref{sec:helicity_offshell} and Sec.~\ref{sec:helicity_scatt}, vastly different behavior is possible outside this limit. Therefore, to further explore off-shell effects, we consider again  the $1\to4$-body decays in Eqs.~\eqref{eq:decay_proc_lnc} and \eqref{eq:decay_proc_lnv} along with the decay asymmetry measure $\mathcal{A}_\Gamma$ in Eq.~\eqref{eq:def_asymm_decay}. We also  consider the $2\to4$-body scattering processes in Eqs.~\eqref{eq:scatt_proc_lnc} and \eqref{eq:scatt_proc_lnv} along with the cross section asymmetry measure $\mathcal{A}_\sigma$ in Eq.~\eqref{eq:def_asymm_xsec}. 

In Fig.~\ref{fig:lnvHelicity_SM_WL_WidthAsymm_LargeWidth_LoMass} we plot the width asymmetry for heavy neutrino masses in the range of $m_N\in[1\GeV, 95\GeV]$  and in Fig.~\ref{fig:lnvHelicity_SM_WL_WidthAsymm_LargeWidth_HiMass} for the larger range $m_N\in[1\GeV,  250\GeV]$. While we set the relevant active-sterile mixing to be $\vert V_{e4}\vert^{2}=10^{-2}$ with $\vert V_{\mu 4}\vert^2, \vert V_{\tau 4}\vert^2 = 0$, we fix $N$'s total width to be $\Gamma_N= 100\MeV$ (dash-dot) and $100\GeV$ (dash). As a reference we consider as well the width according the Lagrangi an in Sec.~\ref{sec:theory_model} (default, solid). The statical MC uncertainty band is also shown.

For neutrino masses below $M_W$ and for both $\Gamma_N= 100\MeV$ and $10\GeV$, we observe the presence of  large, positive  asymmetries. These indicate the enhancement of the $\Delta L = 0 $ decay mode over the $\vert \Delta L \vert=2$ mode, and in comparison to the benchmark case can be attributed to large-width effects. To understand this in terms of kinematics, note that in the absence of a resonance pole the natural scale for $N$'s virtuality is $p_N^2 \sim M_W^2 \gtrsim m_N^2$. The enlarged widths allow $N$'s virtuality to vary away from $p_N^2\sim m_N^2$, which then drives a positive asymmetry according to the analytic expression of Eq.~\eqref{eq:asymm_offshell_largeWidth}.

 We find that $\mathcal{A}_{\Gamma}$ approaches unity for both large-width cases when $m_N\lesssim10\GeV$, and reduces to about $\mathcal{A}_{\Gamma}\sim0.5$, for $\Gamma_N=100\MeV~(10\GeV)$ when $m_N\sim 10~(25)\GeV$. For masses above $m_N\sim45\GeV$ but still below $M_W$, the width-to-mass ratio for $\Gamma_N=100\MeV$ is small enough that off-shell effects become negligible and the asymmetry vanishes. For $\Gamma_N=10\GeV$ and for all values of $m_N <  M_W$, the asymmetry is always greater than the MC uncertainty band, implying that off-shell effects are always significant. At around $m_N\sim75\GeV$, i.e., just below the $W$ boson's mass threshold, the asymmetries for the benchmark and $\Gamma_N=100\GeV$ curves begin to move from $\mathcal{A}_\Gamma\sim0$ to sub-zero values.
 
 For neutrino masses above $M_W$ and for all three width cases, we observe the presence of negative asymmetries that slowly approach $\mathcal{A}_\Gamma=-1$ for increasing $m_N$. This implies a suppression of the $\Delta L = 0$ decay mode over the $\vert \Delta L \vert=2$ mode, and can be attributed to the faster decoupling of ultra heavy $N$ in the  $\Delta L = 0$ ME. (For related details, see the discussion following Eq.~\eqref{eq:asymm_offshell_largeMass}.) At values of $m_N$ just above $M_W$ we find that the width-to-mass ratios for all three cases are negligible enough that all three curves converge. This essentially affirms an insensitivity to the total width in the high-mass limit.
 
Turning to potential impact of off-shell effects in $2\to4$ scattering, we plot in Figs.~\ref{fig:lnvHelicity_SM_WL_XSecAsymm_LargeWidth_LoMass} and \ref{fig:lnvHelicity_SM_WL_XSecAsymm_LargeWidth_HiMass}, respectively, the cross section asymmetry measure for heavy neutrino masses in the range of $m_N\in[1\GeV, 95\GeV]$  and $m_N\in[1\GeV, 250\GeV]$. We assume the same inputs as in Figs.~\ref{fig:lnvHelicity_SM_WL_WidthAsymm_LargeWidth_LoMass} and ~\ref{fig:lnvHelicity_SM_WL_WidthAsymm_LargeWidth_HiMass} for the decay asymmetry.

For lighter states with $m_N < M_W$ we observe that the default and $\Gamma_N=100\MeV$ cases exhibit similar qualitative and quantitative behavior as for the decay asymmetry. In these two cases the total widths for $N$ and $W$ are small and satisfy the relation $\Gamma_N \ll \Gamma_W \ll m_N \lesssim M_W$. The virtuality carried by the $W$ is therefore comparable to its mass, i.e., $Q\sim M_W$, resulting in similar kinematics as in $W$ boson decays. Arguably, for such small $\Gamma_N$, one can treat the decays of $W$ bosons with the spin-correlated NWA. When $\Gamma_N=10\GeV$ we observe a cross section asymmetry that is positive and larger than in the decay case. Quantitatively, the asymmetry reaches approximately $\mathcal{A}_{\sigma}\sim0.75~(0.5)$ when $m_N\sim25~(35)\GeV$, never drops below $\mathcal{A}_{\sigma}\sim0.15-0.2$  for $m_N < M_W$, and briefly grows in the vicinity of $m_N\sim M_W$. The significant differences between the cross section and decay asymmetries at large $N$ widths, particularly when $\Gamma_N\lesssim m_N\lesssim M_W$, demonstrates a breakdown of the NWA.
 
When Majorana neutrinos are heavier but still kinematically accessible, i.e., when $M_W < m_N \ll \sqrt{s}$, the cross section asymmetry rapidly shrinks for each choice of $\Gamma_N$. More specifically, the asymmetry drops to $\mathcal{A}_{\sigma}\sim0.01$ when $m_N\sim35~(500)\GeV$ and $\Gamma_N=100\MeV~(10\GeV)$, and continues toward zero for larger $m_N$. $\vert\mathcal{A}_{\sigma}\vert$ remains small, if not negligible, until $m_N\sim2-5\TeV$. At these masses the cross section asymmetries migrate to negative values of $\mathcal{A}_{\sigma}$, again indicating a $L$-violating cross section that is larger than the analogous $L$-conserving rate. We observe a hierarchy in this behavior, with larger $\Gamma_N$ leading to more negative values of $\mathcal{A}_{\sigma}$. (Note that due to the opening of new decay modes, $\Gamma_N^{\rm default}\gg 10\GeV$ at such large values of $m_N$.)

When $N$ is no long kinematically accessible, i.e., when $M_W \ll \sqrt{s} \lesssim m_N$, one again enters the decoupling phase. Here we observe observe similar behavior as found in the $m_N > M_W$ regime of the decay asymmetry in Fig.~\ref{fig:lnvHelicity_SM_WL_WidthAsymm_LargeWidth_HiMass}. At these scales, the total widths of $N$ are irrelevant as it is always far off-shell. As such, all three curves converge at around $m_N\sim 10\TeV$ and tend toward $\mathcal{A}_{\sigma}=-1$ for increasing heavy neutrino masses. Using again the language of effective field theories, the negative-valued cross section asymmetry follows from the $L$-violating process occurring at dimension seven, whereas the $L$-conserving process occurs at dimension eight. The latter rate is thus relatively suppressed by a factor of $(Q/m_N)^2\ll1$.

\section{Summary and Conclusions}\label{sec:conclude}

Whether or not light neutrinos are Majorana fermions remains one of the most pressing open questions in particle physics today. If neutrinos are their own antiparticle, then it is likely that new particles and interactions play a role in generating neutrino masses that are hierarchically smaller than the EW scale. Hence, establishing the Majorana nature of neutrinos is a stepping stone to more fully understanding the fundamental symmetries of nature.

In this study, we report an analytical and numerical investigation into the impact of helicity inversion on partial widths and cross sections of $\vert\Delta L\vert =  2$ processes at the LHC. We focus as a case study on $L$-conserving and $L$-violating, $4$-body decays of $W$ bosons mediated by a heavy Majorana neutrino $N$ in the Phenomenological Type I Seesaw model. After isolating the relative helicity preservation (inversion) in the $L$-conserving (violating) process at the ME level in Sec.~\ref{sec:helicity_lnc} (\ref{sec:helicity_lnv}), we show that up to finite-width effects an identical dependence on $N$'s mass $(m_N)$ emerges at the squared ME level due to the different scaling of 4-momenta  and squared 4-momenta. When $N$ goes on-shell, we find that this mass dependence precisely cancels. This renders total decay and scattering rates equal and non-zero, even for small $m_N$. We go on to find in Sec.~\ref{sec:helicity_offshell} that for far off-shell $N$, large differences between the  $\Delta L = 0$ and $\vert \Delta L\vert = 2$ squared MEs can arise. Depending on the precise off-shell limit, this can lead to a relative enhancement or suppression of the $L$-violating channel. In Sec.~\ref{sec:helicity_scatt}, we show that our findings extend to $2\to4$ scattering, and in Sec.~\ref{sec:helicity_other} to other neutrino mass models, so long as consistent propagation of helicity inversion is taken into account.

In Sec.~\ref{sec:results} we quantify our analytic results by performing exact numerical ME computations using the MC event generator \texttt{MadGraph5\_aMC@NLO} in conjunction with the \texttt{HeavyN} model libraries. Starting in Sec.~\ref{sec:results_ratio}, we confirm the strong presence of helicity inversion in the $W\to Ne$ decay process  using  our MC setup. We  then move onto  the more general $1\to4$ decay  and $2\to4$ scattering  processes in Secs.~\ref{sec:results_width}  and \ref{sec:results_xsec}, respectively. After building asymmetries $(\mathcal{A})$ sensitive to helicity suppression and enhancements in $L$-violating processes, we report the absence of numerically  significant  helicity suppression when decay and scattering processes are dominated by an on-shell $N$ with a small width. This is despite the presence of helicity inversion. In Sec.~\ref{sec:results_offshell} we find that far off-shell contributions can lead to numerically significant helicity enhancement or suppression of $\vert \Delta L\vert = 2$ channels, particularly in the decoupling limit,  as well as  to a breakdown of the NWA. In all cases, we find strong agreement with analytical expectations for on- and off-shell Majorana neutrinos.

Taking everything together, we condense our findings into the following model-dependent conclusions:
\begin{itemize}
\item When a Majorana neutrino $N$ can be on-shell and its width $\Gamma_N$ is small, i.e., $\Gamma_N\ll m_N$, then $L$-conserving and $L$-violating rates are the same. 
\item When $N$ can be on-shell but its width is large, i.e., $\Gamma_N\sim m_N$, then off-shell/finite-width effects trigger $L$-conserving rates larger than $L$-violating rates.
\item When $N$ is too heavy  to be on-shell, e.g., $m_N \gtrsim \sqrt{s}$,  then off-shell/decoupling effects trigger $L$-violating rates that are larger than $L$-conserving rates.
\end{itemize}
We stress that more could be learned by further investigating finite width effects as well as the criteria for applying the NWA when Majorana fermions are present. More can also be learned about the potential loop-level generation of helicity asymmetries $\mathcal{A}$ as well as the role of additional new particles with chiral interactions. We strongly encourage future studies.

\section*{Acknowledgements}
Didar Dobur and Lesya Shchutska are acknowledge for discussions that initiated this work. Andr\'e de Gouv\^ea and Boris Kayser are thanked for their comments on an early version of this manuscript. Carlos Alishaan, Alexey Boyarsky, Albert De Roeck, and Olivier Mattelaer are also thanked for their helpful input.

This work has received funding from the European Union's Horizon 2020 research and innovation programme as part of the Marie Skłodowska-Curie Innovative Training Network MCnetITN3 (grant agreement no. 722104), 
FNRS “Excellence of Science” EOS be.h Project No. 30820817.
This work is also supported under the UCLouvain fund “MOVE-IN Louvain.”
Computational resources have been provided by the supercomputing facilities of the Universit\'e catholique de Louvain (CISM/UCL)  and the Consortium des \'Equipements de Calcul Intensif en F\'ed\'eration Wallonie Bruxelles (C\'ECI) funded by the Fond de la Recherche Scientifique de Belgique 
(F.R.S.-FNRS) under convention 2.5020.11 and by the Walloon Region.


\end{document}